\documentclass{article}
\usepackage[utf8]{inputenc}

\usepackage{fancyhdr}
\usepackage{listings}
\usepackage{graphicx}
\usepackage{color}
\usepackage{amsmath}
\usepackage{amssymb}
\usepackage{setspace}
\usepackage{authblk}
\lstdefinestyle{mystyle}{
    commentstyle=\color{blue},
    keywordstyle=\color{magenta},
    numberstyle=\tiny\color{codegray},
    basicstyle=\footnotesize,
    breakatwhitespace=false,         
    breaklines=true,                 
    captionpos=b,                    
    keepspaces=true,                 
    numbersep=5pt,                  
    showspaces=false,                
    showstringspaces=false,
    showtabs=false,                  
    tabsize=2
}
 
\title{Using data-reduction techniques to analyse biomolecular trajectories}
\author[a]{Gareth A. Tribello\thanks{corresponding author: g.tribello@qub.ac.uk}}
\author[b]{Piero Gasparotto}
\affil[a]{Atomistic Simulation Centre, School of Mathematics and Physics, Queen's University Belfast, Belfast, BT7 1NN, United Kingdom}
\affil[b]{Laboratory of Computational Science and Modelling and National Centre for Computational Design and Discovery of Novel Materials MARVEL, IMX, {\'E}cole Polytechnique F{\'e}d{\'e}rale de Lausanne, 1015 Lausanne, Switzerland}

\begin{document}

\maketitle

\begin{quotation}
This chapter discusses the way in which dimensionality reduction algorithms such as diffusion maps and sketch-map can be used to analyze molecular dynamics trajectories.  The first part discusses how these various algorithms function as well as practical issues such as landmark selection and how these algorithms can be used when the data to be analyzed comes from enhanced sampling trajectories.  In the later parts a comparison between the results obtained by applying various algorithms to two sets of sample data are performed and discussed.  This section is then followed by a summary of how one algorithm in particular, sketch-map, has been applied to a range of problems.  The chapter concludes with a discussion on the directions that we believe this field is currently moving.     
\end{quotation}

\begin{quotation}
molecular dynamics $\vert$ dimensionality reduction $\vert$ sketch-map 
\end{quotation}

\section{Introduction}

The first molecular dynamics (MD) simulation of a biomolecule was performed in 1977 \cite{bovine-protein}. The 9.2~ps trajectory for the bovine pancreatic trypsin inhibitor that was extracted from this work and the countless longer simulations that have followed have fundamentally changed our view of biomolecules.  We now no longer believe that proteins, DNA and so on are simply rigid structures and instead acknowledge that the dynamical motions of these molecules are often critical to their functions.  Dynamical simulations are thus an essential tool when it comes to the study of these complex structures.  The problem, however, is that the trajectories that emerge from these studies contain almost too much information as they describe how the positions and velocities of all the atoms within the protein change as a function of time.  On top of this biomolecules, unlike simpler systems, have energy landscapes that are very complicated.  Consequently, unlike crystalline solids or clusters of indistinguishable atoms, biomolecules do not normally undergo transitions that involve a change in symmetry.   It is therefore difficult to find the lowest energy configuration for a biomolecule and to develop a rationale for analyzing the results from a simulation \cite{energy-landscapes}.  

The lack of a simple theoretical framework based on symmetry for rationalising the behavior of all biomolecules together with the abundance of dynamical information that can be easily extracted by performing long molecular dynamics simulations has led many researchers to use machine learning algorithms when analysing trajectories of biomolecules.  In this chapter we will document some of this work.  Before discussing algorithms, however, we first note that it is important to think carefully about what information emerges when these machine learning algorithms are used to analyze molecular dynamics trajectories.  In essence all the algorithms we will discuss in this chapter treat the trajectory as a set of high dimensional vectors.  They then generate a representation of this data  that definitely has a lower information content by attempting to capture the most important features from the input data.  In other words, all the algorithms we will discuss perform a data reduction operation on a set of high-dimensional vectors.  In practice, this data-reduction operation is achieved by either: 
\begin{enumerate}
    \item Selecting a small number of representative points in this high dimensional space and asserting that the variations between this small number of points describe all the important variations between the points in the larger input data set. 
    
    \item Generating projections of each of the high-dimensional vectors in some lower dimensional space and assuming that the variation between the structures that is observed in the high dimensional space can be represented in this lower dimensional space. 
\end{enumerate}
All the algorithms that are described in this chapter adopt one of the above strategies or both strategies in combination.  The critical thing to remember, however, with both of these approaches and, by extension about the algorithms that will be discussed in this chapter, is that a mathematical model is used to reduce the high-dimensional data.  Using this model introduces assumptions about the structure of the data as models cannot be fitted to data without making assumptions.  Hence, when using method (1), we assume that most of the trajectory frames will be clustered about one or more representative structures.  This assumption seems reasonable given that we know from statistical mechanics that the system will, at low temperature, for the most part remain close to the deepest basins in the potential energy landscape.  We should not forget the so-called curse of dimensionality, however, and the fact that many of the very-standard algorithms that we might use to analyse low-dimensional data cannot necessarily be applied to high-dimensional data \cite{curse-of-dim}.  

The theoretical case for using method (2) is much less well established.  Studies have shown that the dimension of the space explored by a protein containing $N$ atoms is considerably lower than $3N$, which is what would be expected based on the the number of degrees of freedom \cite{pca-traj-1,pca-traj-2,pca-traj-3,how_complex_proteins,new-laio-dimension,more-clementi-and-noe}.  Furthermore, the fact that biomolecules behave in predictable ways suggests that the potential constrains these molecules to explore only a small fraction of configuration space.  Importantly, however, the assumption in many data reduction algorithms of type (2) is that the biomolecule can only adopt configurations that lie on a low dimensional linear or non-linear manifold.  This assumption is considerably stronger than the assertion that system is confined in a small region of configuration space.  The system might, for instance, be confined in a region with a low fractal dimension, which cannot be represented in a low dimensional Euclidean space \cite{fractal-manifold-1}.  

This brings us to the most critical piece of advice we would give to a person just starting out with these algorithms.  The techniques described in this chapter are tools for visualizing the data contained in a trajectory.  It is very important to understand how they function and what assumptions they make.  Most critically, however, these algorithms are not a replacement for chemical or physical intuition.  When used well though they may enhance our ability to make leaps in understanding by clearing away distractions. 

\section{Theory}

There are many excellent books, papers and online resources on machine learning that cover the wider theory of these dimensionality reduction algorithms \cite{MDS_book,PCA_book,stats_book}.  What we will thus do here is provide a  summary of the important considerations that must be taken into account when using these algorithms to analyze trajectory data.  In particular, we will discuss how the data should be collected in section \ref{sec:collect}. We will then talk about what is input into these algorithms in section \ref{sec:represent} and how a subset of so-called landmark points can be selected in section \ref{sec:landmarks} before describing how the algorithms operate in section \ref{sec:dim-red}.  
In addition, to presenting this material we will show a number of case studies that demonstrate the way these algorithms have been used to analyze trajectories of biomolecules.  We will then finish by speculating on the general direction in which we believe the field is moving.  

\subsection{Step 1: Collecting some data}
\label{sec:collect}

Obviously, we cannot perform any form of machine learning without first collecting some data.  The first thing we must then consider when performing any form of machine learning is the manner in which the data is collected.  For the purposes of this chapter we will assume that data has been generated by performing a molecular dynamics simulation \cite{frenkel-schmidt,all-tild90book}.  As has been discussed at length in many of the other chapters in this book, however, collecting representative data on the regions of configuration space that a biomolecule typically samples from in this way is not straightforward.  The problem is that there are often large barriers that prevent the molecule from diffusing freely around all of configuration space.  These barriers are typically not crossed during short molecular dynamics simulations and thus configuration space is only partially sampled.  The consequence of this when we run our machine learning algorithm on the generated data is that the simplified view that we construct only gives a partial insight into the structure of configuration space.  There is, obviously, no guarantee that any representation extracted from this data gives a reasonable representation for the parts of configuration space that were not sampled during the molecular dynamics trajectory.

Other chapters in this book discuss a range of enhanced-sampling algorithms that allow us to resolve this timescale problem.  When these algorithms are used we change the manner in which the system samples configuration space by either adding a simulation bias to the potential or by adding new ways for the system to move around in configuration space.  Furthermore, when these techniques are used free energy differences and barrier heights for the unbiased ensemble can be extracted by exploiting reweighting techniques.  These reweighting methods are important in the context of machine learning as they should be used when data generated using an enhanced sampling technique is analyzed using a machine learning algorithm. 

\begin{figure}
    \centering
    \includegraphics[width=0.8\textwidth]{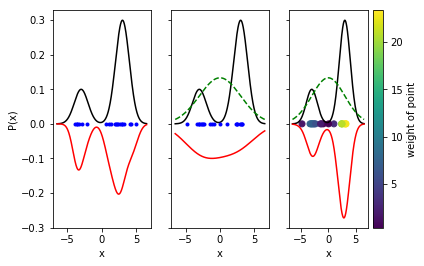}
    \caption{Figure illustrating how we can use reweighting algorithms to extract information on the unbiased distribution from a biased trajectory.  The black line in the right panel shows a probability distribution and a set blue dots that represent 20 samples that we have generated from this distribution.  The red line then shows the estimate of the probability density that we extract when we perform a kernel density estimation using this data.  The middle panel shows something similar but this time we have generated our blue samples from the distribution shown as a dashed green line.  Consequently, the estimate of the probability density that we get by performing a kernel density estimation using this data (red line) resembles the green line and not the black line.  In the third panel, however, we show that if we ascribe a weight to each of the points sampled from the green distribution using the formula in the text we can recover a probability density function from kernel density estimation (red line) that resembles the black curve.  The substantial differences between the underlying distributions and the estimates we get from kernel density estimation are due to the limited sampling, which is something that would also need to be considered when analyzing trajectory data. }
    \label{fig:reweighting}
\end{figure}

Ultimately, we know from elementary statistical mechanics that, if a protein is at equilibrium, we can think of the configuration, $\mathbf{X}$, that it adopts at any given instance in time as a random vector taken from some high dimensional probability distribution $P(\mathbf{X})$ that depends on the macroscopic state, which in this case means that it depends on the number of atoms, $N$, the volume, $V$ and the temperature, $T$.  We can thus think of the configurations sampled during an unbiased molecular dynamics simulation as a series of random vectors, $\{\mathbf{X}_i\}$, that are generated from $P(\mathbf{X})$.  Furthermore, when we analyze an unbiased molecular dynamics simulation using a machine learning algorithm we exploit the law of large numbers and the central limit theorem and assume that the distribution of sampled vectors  provides us information on this probability distribution $P(\mathbf{X})$.  When we use an enhanced sampling algorithm, however, the configurations sampled can no longer be thought of as random vectors generated from $P(\mathbf{X})$.  The problem with doing so being that to achieve the greater rates of sampling we changed the Hamiltonian or the thermodynamic constraints.  The configurations generated from these enhanced sampling trajectories are thus samples from some other probability distribution, $P'(\mathbf{X})$.  All is not lost, however, because, as discussed in the other chapters of this book, there are simple recipes from extracting information on $P(\mathbf{X})$ from a set of samples of $P'(\mathbf{X})$.  The way these methods work is illustrated in figure \ref{fig:reweighting}.  To generate the right panel of this figure we generated 20 random variables from the probability distribution, $P(x)$, that is shown using a black line.  The values of these random variables are indicated using blue dots on the x axis of the figure.  A kernel density estimation was then performed using these points as input in order to generate the estimate of the probability density function that is shown inverted and in red in the figure.  To generate the middle panel we instead generated points using the probability distribution, $P'(x)$, that is shown in green in the central panel of the figure.  The 20 random variables generated from this distribution are once again shown in green and you can see that the estimate of the probability density that we construct by performing a kernel density estimation using this data bears no resemblance to the black line, $P(x)$, and instead resembles the green dashed line, $P'(x)$.  The third panel of the figure shows, however, that if we ascribe a weight:
$$
w_i = \frac{P(x_i)}{P'(x_i)}
$$
to each of the points generated, $x_i$, we can recover the probability distribution $P(x)$ even if we sample from $P'(x)$.  In this final panel the points sampled from $P'(x)$ are shown on the $x$-axis once more and are coloured according to the value of $w_i$.  The estimate of the probability density function that is shown in red is then calculated using:
$$
\sum_{i=1}^N w_i K( x_i )
$$
where the sum runs over the number of data points generated and where $K$ is a Gaussian kernel.  As you can see this estimated probability density function in this final panel resembles the black line, $P(x)$.  Consequently, if we analyze appropriately-weighted  configurations using a machine learning algorithm the representation that is extracted provides information on $P(\mathbf{X})$.  In the remainder of this chapter we will thus assume that the input to our machine learning algorithm consist of a set of high-dimensional vectors, $\{\mathbf{X}_i\}$, and a set of associated weights $\{w_i\}$.  If the data to be analyzed comes from an unbiased molecular dynamics trajectories these weights are all set equal to one.  We need to have this flexibility to give the vectors different weights, however, in order to deal effectively with data from enhanced sampling trajectories.  

\subsection{Step 2: Representing the data}
\label{sec:represent}

In the previous section we discussed the collection of data from biased and unbiased molecular dynamics trajectories in abstract terms.  The trajectories output from these methods were thought of as a set of random high-dimensional vectors with associated weights.  In this section we will discuss more precisely what information we might want to collect from these vectors.  The key point is that we want to throw away information that we know is irrelevant at an early stage as otherwise any interesting signal that we might detect with a machine learning algorithm will be lost in a sea of noise.  As an example, if we were simulating the dynamics of a protein in water, we could simply collect the positions of all the protein and water atoms in the system.  This is probably self defeating, however, as the number of atoms of water outnumbers the number of protein atoms by far and our interest is in the behavior of the protein and not the water.  With this in mind we should thus probably only collect information on the positions of the protein atoms.  Even this might be more than we require, however.  The trajectory for all the protein atoms probably contains a significant amount of noise that describes the small fluctuations of the atomic positions around equilibrium positions whereas we are probably more interested in larger-scale, global motions that result in a significant change in the protein's conformation.  We might, therefore, be tempted to throw away most of this information on the atomic positions and to instead collect only the values of the backbone torsional angles.

The point we are trying to make is this: you shouldn't disregard you physical or chemical intuition just because you are using a machine learning algorithm.  In other words, these algorithms should be used to complement your intuition about the system in question and not to replace it.  The plain fact is that you are more likely to get an informative projection of your trajectories if you use what you know to ensure that there is not too much noise in the input data.

\begin{figure}
    \centering
    \includegraphics{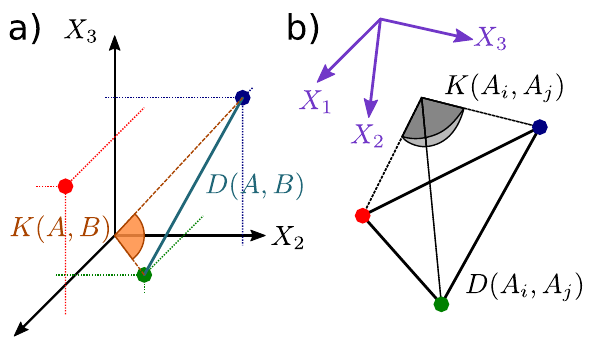}
    \caption{Figure illustrating the three possible representations of the data contained in a trajectory that can be used as the input in these dimensionality reduction algorithms.  The data can either be represented as a set of fingerprint vectors, $X$, that describe the positions of the trajectory frames in some feature space.  Alternatively, the dissimilarity, $D$, between each pair of configurations can be computed and stored in a matrix.  Lastly, the inner product, $K$, between each pair of fingerprints can be computed and these quantities can be stored in a Gram matrix.}
    \label{fig:representation}
\end{figure}

Figure \ref{fig:representation} illustrates a further consideration that is important when it comes to the representation of the trajectory.  Remember that when we use a machine learning algorithm to analyze a trajectory, $\{ \mathbf{X}_i \}$, what we are essentially trying to illustrate is how these random vectors are distributed in relation to each other.  There are, however, three different ways that we can use to illustrate these relations in the high dimensional space.  In particular:

\begin{enumerate}
\item We can use a vector, $\mathbf{X}_i$, of fingerprints to represent each of the configurations.  The various components of this vector represent the projections of the vector connecting the origin and the point $\mathbf{X}_i$ on some arbitrarily chosen axes.

\item We can use a dissimilarity matrix, $\mathbf{D}$ in which element $D_{ij}$ gives the distance between configuration $i$ and configuration $j$.  This distance can be calculated using any metric.

\item We can use a Gram matrix, $\mathbf{K}$ in which element $K_{ij}$ gives the inner product between the vectors of fingerprints for configurations $i$ and $j$. 
\end{enumerate}

It is straightforward to convert between these three different ways of representing the data.  For instance, if you are given vectors of fingerprints you can clearly compute the matrices of inner products, $\mathbf{K}$, or the matrix of distances, $\mathbf{D}$.  What is perhaps less obvious is that you can compute a matrix of inner products, $\mathbf{K}$, from a matrix of distances, $\mathbf{D}$, and that you can convert any matrix of inner products into a set of vector fingerprints.  To convert the matrix of distances into a matrix of inner products we exploit the fact that we can write the $(i,j)$-element of the matrix of squared distances as follows:
\begin{equation}
D_{ij}^2 = \sum_\alpha ( \mathbf{X}^{(i)}_\alpha - \mathbf{X}^{(j)}_\alpha)^2 = \sum_\alpha (\mathbf{X}^{(i)}_\alpha)^2 + (\mathbf{X}^{(j)}_\alpha)^2 - 2 \mathbf{X}^{(i)}_\alpha \mathbf{X}^{(i)}_\alpha
\label{eqn:pythagoras}
\end{equation}
where the symbol $\mathbf{X}_\alpha^{(i)}$ is used to represent the $\alpha$th component of the vector of fingerprints for configuration $i$. 
Notice that the final term in this expression is the $(i,j)$-element of the Gram matrix of dot products, $\mathbf{K}$, that we require.  Furthermore, the first and second terms are independent of $j$ and $i$ respectively.  
We can thus rewrite the matrix of dissimilarities $\mathbf{D}$ as (see Note \ref{note:distance-matrix}):
\begin{equation}
\mathbf{D} = \mathbf{c}\mathbf{1}^T + \mathbf{c}^T \mathbf{1} - 2 \mathbf{K} 
\label{eqn:dissims-matrices}
\end{equation}
In this expression $\mathbf{c}$ and $\mathbf{1}$ are column vectors containing the same number of elements as there are rows in $\mathbf{D}$.  All the elements of $\mathbf{1}$ are equal to 1 and the $i$th element of $\mathbf{c}$ is equal to $\sum_\alpha (X_{\alpha}^{(i)})^2$.  We now introduce the centering matrix:
\begin{equation}
\mathbf{J} = \mathbf{I} - \frac{1}{M} \mathbf{1} \mathbf{1}^T
\label{eqn:centering-matrix}
\end{equation}
where $\mathbf{I}$ is the identity and $M$ is the number of rows in $\mathbf{D}$.  This centering matrix is useful because if $\mathbf{D}$ is multiplied from the front and back by $-\frac{1}{2} \mathbf{J}$ we recover the Gram matrix of kernels modulo an additive constant (see Note \ref{note:centering}).

Extracting vectors of fingerprints from a Gram matrix is similarly straightforward.  To do so we begin by considering a rectangular matrix, $\mathbf{X}$ that contains the $i$th vector of fingerprints in its $i$th column.  The Gram matrix can be calculated from $\mathbf{X}$ by computing the following product of matrices:
\begin{equation}
\mathbf{K} = \mathbf{X}^T\mathbf{X}
\label{eqn:gram}
\end{equation}
The matrix $\mathbf{K}$ that we compute in this way is symmetric and has all real elements so we can thus exploit the spectral decomposition for symmetric matrices and write:
\begin{equation}
\mathbf{K} = \mathbf{V} \Lambda \mathbf{V}^T
\label{eqn:spectral}
\end{equation}
where $\Lambda$ is a diagonal matrix containing the eigenvalues of $\mathbf{K}$ and where $\mathbf{V}$ is a matrix containing the corresponding eigenvectors in its columns.  Comparing equations \ref{eqn:gram} and \ref{eqn:spectral} we thus find that the matrix of fingerprints is given by:
$$
\mathbf{X}^T = \mathbf{V}\Lambda^\frac{1}{2}
$$
In other words, the eigenvectors of the Gram matrix can serve as a basis on which we can project each of our configurations.  Furthermore, this process of collecting fingerprints, calculating a matrix of dissimilarities between them using Pythagoras' theorem, centering this matrix and then diagonalizing it can be seen as equivalent to principal component analysis in that it is simply a rotation of the reference frame on which we are projecting our fingerprint vectors.

The fact that we can prove that these three ways of representing the high-dimensional data are all equivalent may feel like a pointless exercise in linear algebra.  After all, you may ask, wouldn't we always collect a vector of fingerprints from the trajectory?  In other words, are we not always computing $\mathbf{D}$ and $\mathbf{K}$ from fingerprint vectors?  The answer to this is yes but in an equally real sense no as, although we do always collect vectors of data, we may choose to not calculate the matrix of dissimilarities, $\mathbf{D}$, between these vectors by simply using Pythagoras theorem as we did in equation \ref{eqn:pythagoras}.  For example, we might choose to calculate the $(i,j)$ element of $\mathbf{D}_{ij}$ by taking the root mean square deviation between the positions of the atoms in frame $i$ and frame $j$ in a way that removes the motion of the centre of mass and the rotation of the reference frame \cite{rmsd}.  Alternatively, in the manifold learning method known as ISOMAP, the dissimilarities in the matrix $\mathbf{D}$ are representative of the geodesic distances between trajectory frames \cite{isomap,landmark-isomap}.  In both these cases the vectors that emerge are thus no longer related to the vectors that were collected from our trajectory by a rotation.   

An even more interesting case is presented in the method known as kernel principal component analysis \cite{kernel-pca}.  This method was developed based on the observation that, although it may not be possible to linearly separate $N$ points in $d<N$ dimensions, it will almost always be possible to do so in $d \ge N$ dimensions.  This method thus argues that we should thus begin by mapping each of $N$ points, $\mathbf{x_i}$, in our $d<N$ dimensional space into an $N$-dimensional space using some function $\Phi$ such that:
$$
\Phi(\mathbf{x}_i) \qquad \textrm{where} \qquad \Phi : \mathbb{R}^d \rightarrow \mathbb{R}^N
$$
The problem is that we do not know how this mapping should be done in practice.  This difficulty is avoided in kernel-PCA, however, which exploits the so-called kernel trick.  This trick relies on the fact that certain functions, $k$, of pairs of vectors, for example $k(\mathbf{x},\mathbf{y}) = \exp(-|\mathbf{x}-\mathbf{y}|)$, can be expressed as inner products in a high dimensional space.  In other words:
$$
k(\mathbf{x},\mathbf{y}) = \Phi(\mathbf{x})^T\Phi(\mathbf{y})
$$
In practice what this relation means for kernel PCA is that we do not need to determine the mapping $\Phi(\mathbf{x}_i)$.  Instead we can calculate the Gram matrix by evaluating $k(\mathbf{x},\mathbf{y})$ for each pair of configurations from our trajectory.  By diagonalizing the Gram matrix we thus get vector fingerprints, $\Phi(\mathbf{x}_i)$, for each of the configurations in our trajectory.  Once again, however, the set of operations that we perform when we use this method is not equivalent to a rotation of the basis vectors.  What we are doing instead, albeit indirectly, is projecting the data into some higher dimensional space.

In summary two important points have been covered in this section:

\begin{itemize}
\item The high dimensional data collected from a trajectory is often noisy.  Much of this noise is due to thermal fluctuations that are not that interesting, however.  Consequently, only data that is believed to be relevant to the phenomenon should be collected from the trajectory.  Many dimensionality reduction algorithms can deal with noise but if there is a lot of noise in your trajectory it becomes increasingly unlikely that you will see anything interesting when it is analyzed.

\item The high dimensional data collected from a trajectory can be represented using either vectors of fingerprints, a dissimilarity matrix or a Gram matrix.  It is possible to convert between these various representations, which is important because, as we shall see in section \ref{sec:dim-red}, many dimensionality reduction algorithms work by simply converting between these various representations. 
\end{itemize}

\subsection{Step 3: Selecting landmarks}
\label{sec:landmarks}

\begin{figure}
    \centering
    \includegraphics[width=0.8\textwidth]{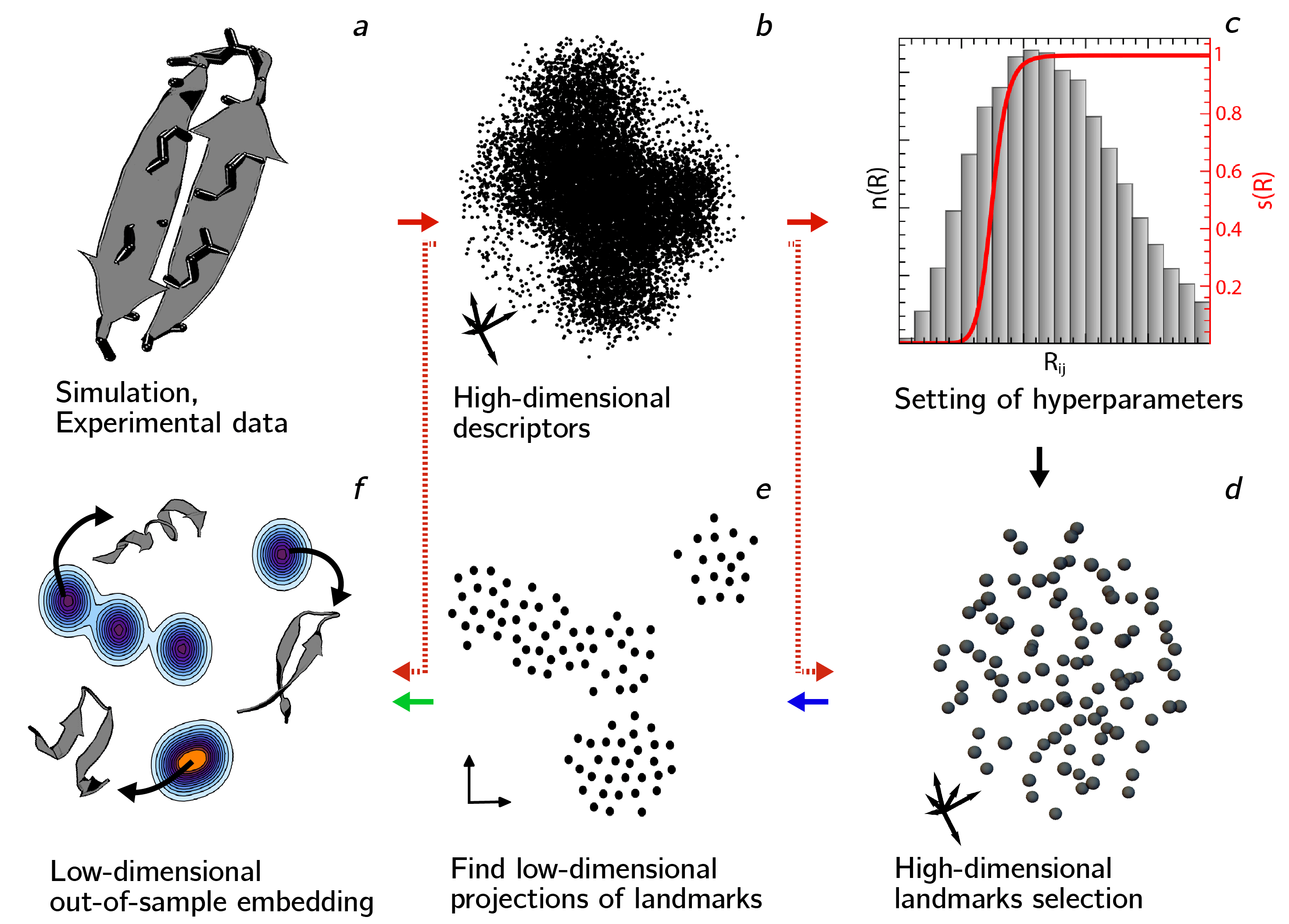}
    \caption{Figure illustrating a workflow that is often used when dimensionality reduction algorithms are used to analyze a trajectory.  In panel (a) of this scheme a trajectory is collected that describes how all the atom positions change as a function of time.  As discussed in the early parts of section \ref{sec:represent} it is often beneficial to calculate a large number of descriptors that describe the processes that you are interested in instead of working with the positions of all the atoms in the trajectory directly.  This is thus what is illustrated in panel (b) above.  Once we have this high dimensional representation we then analyze all the input data in order to set the hyperparameters for the dimensionality reduction algorithm.  Furthermore, as indicated by the red arrow connecting panels (b) and (d) we also select a subset of so-called landmarks points to analyze.  The blue arrow connecting panels d and e indicates that only the landmark points are analyzed using the dimensionality reduction algorithm. Projections for the remainder of the trajectory are found using an out of sample procedure that takes the projections that were found for the landmarks (green arrow) and the high-dimensional descriptions for all the points in the trajectory as input.}
    \label{fig:smap-scheme}
\end{figure}

Many of the algorithms that can be used to analyze trajectory data scale quadratically or cubically with the number of input vectors.  Consequently, these algorithms cannot be used to analyze all the structures in a molecular dynamics trajectory as the associated computational expense would be too large.  One must, therefore, select a small number of representative, landmark structures to analyze using the expensive algorithm.  Furthermore, it is useful to have an out-of-sample algorithm that allows you to construct a representation for any configuration that is outside this initial training set as you can then adopt a work flow like that shown in figure \ref{fig:smap-scheme}. In other words, you can first analyze a small fraction of the input data points using the expensive algorithm and then analyze the remainder of the points using the cheaper out of sample method.

The simplest method for reducing the number of high dimensional vectors from $\{ X_i \}$ that have to be analyzed is to randomly select a smaller number of points from the input data set.  If one is analyzing a molecular dynamics trajectory that has output 10,000 frames this is very straightforward. A random selection of 1,000 vectors can be obtained by simply taking every 10th vector from the larger data set.  Life is made slightly more complex if one is analyzing data from an enhanced sampling trajectory because, as discussed in section \ref{sec:collect}, the weights, $w_i$, associated with the random vectors are no longer all equal to one.  These weights should thus be considered when drawing landmark points as the distribution of landmarks should be consistent with the probability distribution of interest $P(\mathbf{X})$.  Incorporating these weights is not difficult, however \cite{voter-kmc}. The python code below explains how {\it N} points can be drawn from a list of random vectors, {\it R} with weights in a second list, {\it W}, in practice.

\singlespacing
\begin{lstlisting}[caption=Selecting landmarks at random]
def select_random( N, W, R ) :
    # Calculate sum of all weights
    totw = sum( W )
        
    tt, landmarks = 0, []
    for i in range(0,N):
        # Generate a random number between 0 and the total weight 
        # of the unselected points
        tw, rand = 0, (totw-tt)*random.uniform(0,1)
        for j in range(0,len(R)):
            # Make sure each landmark is only selected once
            if R[j] in landmarks : continue
            
            tw += W[j]
            if rand < tw :
                landmarks.append(R[j])
                tt += W[j]
                break

    return landmarks
\end{lstlisting}
\doublespacing

Oftentimes selecting landmark configurations at random is not optimal.  For example suppose that the trajectory samples from a deep basin in the energy landscape and the surrounding, higher-energy regions.  If the landmark points are selected at random they will be distributed in space in a manner that is consistent with the probability distribution $P(\mathbf{X})$.  Consequently, the majority of the selected landmarks will lie inside the basin and very few landmarks from the higher energy regions that surround the basin will be selected.  This selection would not be ideal as any lower dimensional representation generated by analyzing these landmarks may not provide a good description outside of the basin as the algorithm was provided with no data in these regions.  For these reasons a popular alternative to using random sampling of landmarks is to use a method known as farthest point sampling (FPS) \cite{fps}.  As the name suggests this method tries to select the most widely spread set of landmarks from the input random vectors. In other words, the first landmark, $L_1$, is selected at random and the remaining landmarks are selected from the set of all random vectors, $\{R_i\}$, using the following deterministic criteria:
$$
L_{j+1} = \max_{R \in \{R_i\}} \min_{k \le j} |L_k - R |
$$
where $|L_k - R|$ is the dissimilarity between the random vectors $L_k$ and $R$.  A sample python code that performs farthest point sampling is provided below.  In this code it is assumed that the function distance returns the dissimilarity between two random vectors. 

\singlespacing
\begin{lstlisting}[caption=Farthest point sampling]
def farthest_point_sampling( N, W, R ) :
    # Select the first landmark at random
    ll = select_random(1, W, R)
    landmarks = [ ll[0] ]
    for i in range(1,N):
        # The outer loop ensures that the new landmark is the 
        # furthest landmark from the set of landmarks that have 
        # been selected thus far.
        maxd = 0.0
        for rr in R :
            # The inner loop here finds the minimum distance 
            # between data point rr and the set of landmarks that 
            # have been selected thus far.
            mind=float(`Infinity')
            for ll in landmarks :
                if distance(ll,rr)<mind : mind = distance(ll,rr)
        if mind>maxd : 
            maxd = mind
            tland = rr
            
        landmarks.append(tland)
        
    return landmarks
\end{lstlisting}
\doublespacing

\begin{figure}
    \centering
    \includegraphics[width=0.7\textwidth]{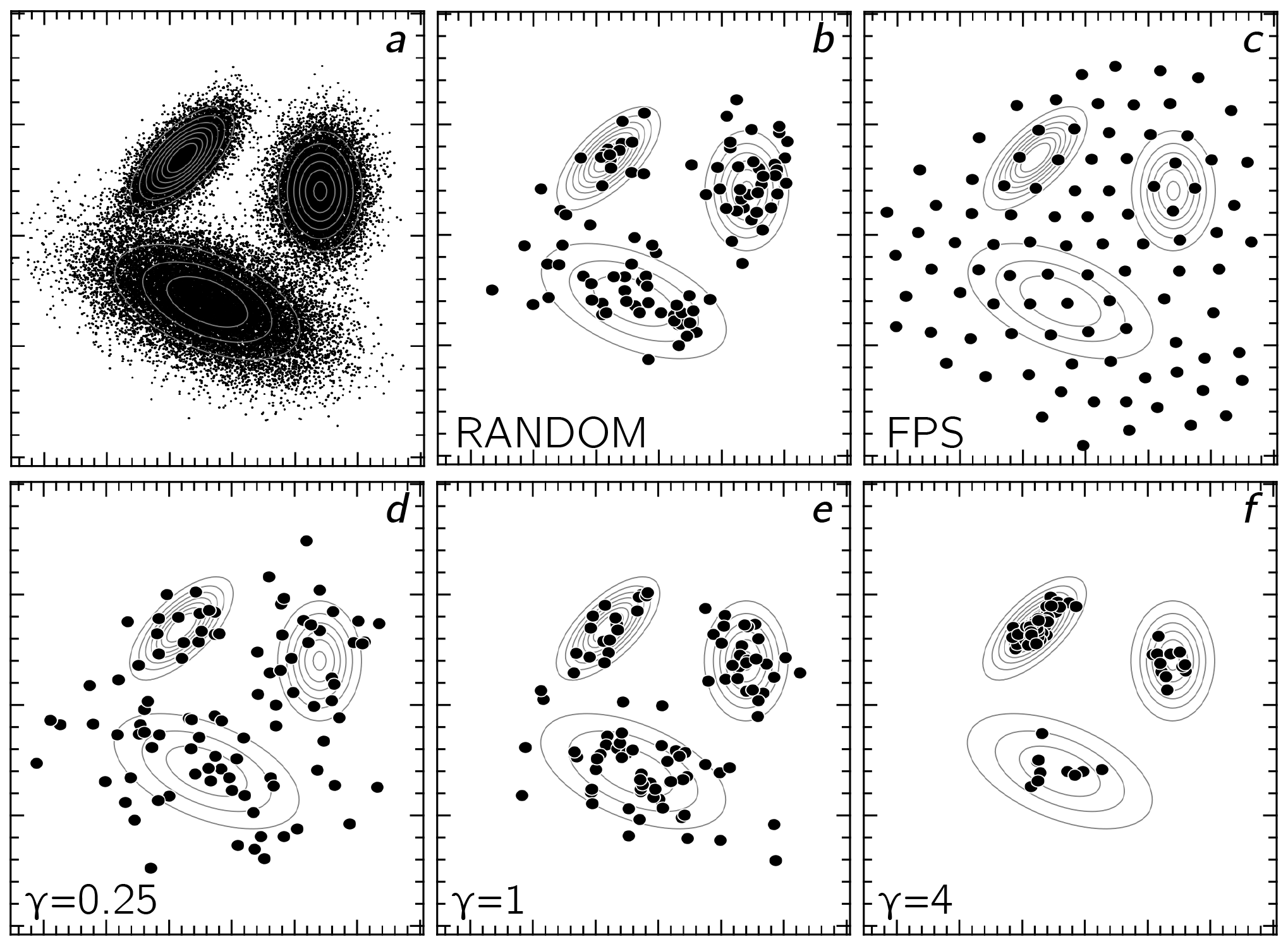}
    \caption{Figure showing how the various landmark selection algorithms perform on model data. Panel (a) shows a set of data points that were generated by sampling from three 2D normal distributions.  The remaining panels then show the set of landmarks that are selected from this data set with each of the algorithms described in the text together with a representation of the three normal distributions that the original data was generated from.  As you can see if random sampling is used the selected landmarks are concentrated in the regions where the density of points is highest.  When FPS is used, by contrast, the points are uniformly distributed across the whole space.  The bottom three panels show that adjusting the $\gamma$ parameter in the well tempered farthest point sampling algorithm allows you to interpolate between these two behaviors and to control the degree to which the points are spread out.}
    \label{fig:staged-algorithm}
\end{figure}

Selecting landmarks using FPS is an improvement on selecting landmarks at random because, as shown in the top part of figure \ref{fig:staged-algorithm}, by using this algorithm we ensure that all the areas of phase space that were sampled during the trajectory are represented in the final set of landmark points.  One disadvantage, however, is that the distribution of landmarks that we get from this procedure no longer provides information on $P(\mathbf{X})$.  We can, however, resolve this problem by giving each of the landmark points, $\{L_j\}$, generated using the FPS algorithm a weight. These weights, $\{\omega_j\}$, can be generated from the weights, $\{w_i\}$ of the input data points, $\{R_i\}$, using a Voronoi diagram as follows:
$$
\omega_j = \sum w_i \quad \textrm{for all vectors in} \quad R_i \quad \textrm{that have} \quad | R_i - L_j | < | R_i - L_k | \quad \forall \quad k \ne j 
$$
A sample python code that calculates the Voronoi weights for the landmarks in the list $L$ from a list containing the input random vectors $R$ and a list containing the weights of those vectors $W$ is provided below.  Notice that this code also calculates the set of random vectors that is in each of the Voronoi polyhedra and that, as in the previous code, the function distance returns the dissimilarity between two random vectors.

\singlespacing
\begin{lstlisting}[caption=Calculating Voronoi weights]
def voronoi_weights( L, R, W ) :
    weights = [0]*len(L)
    points = [[]]*len(L)
    # Loop over all random vectors in data set
    for i in range(0,len(R)) :
        nearest, mind = 0, distance( L[0], R[i] )
        # Find closest landmark to ith random vector
        for j in range(1,len(L)) :
            dist = distance( L[j], R[i] )
            if( dist< mind ) : 
                mind = dist
                nearest = j

        # Add weight of ith random vector 
        # to weight of closest landmark 
        weight[nearest] += W[i]
        # Also add the ith random vector to the list of 
        # random vectors that are assigned to this landmark
        points[nearest].append(R[i])

    return weights, points
\end{lstlisting}
\doublespacing

A slight concern when using FPS sampling to draw landmarks is that the algorithm is rather sensitive to outliers.  To resolve this problem we thus developed a procedure that combines the strengths of FPS and random sampling of landmarks and that involves a two-stage selection process \cite{smap-robust}.  When this procedure is used to select $M$ landmarks from a set of $N$ random vectors the first of these stages involves selecting $K=\sqrt{NM}$ vectors using farthest point sampling.  The top right panel in figure~\ref{fig:staged-algorithm} demonstrates that it is reasonable to assume that these  points distributed uniformly across the space so we can further assume that all the Voronoi polyhedra have the same volume and that the quantity:
\begin{equation}
P_i = \frac{ \omega_i }{\sum_{j=1}^{K} \omega_j }
\label{eqn:vweights}
\end{equation}
thus provides a measure of the probability density in the vicinity of the center of the polyhedron.  In this expression $\omega_i$ is the weight of the $i$th landmark selected, which is calculated from the weights of the data points that were input $\{w_i\}$ using the Voronoi procedure that was outlined in the previous paragraph.  It is interesting to note that, if we now select $M$ points by first picking a Voronoi polyhedron by performing a random sampling using the weights of the polyhedra and if we then select one of the random vectors that is within that Voronoi polyhedron at random, we recover the random sampling method albeit via a rather convoluted route.  More intriguingly, however, we can modify the weights calculated using equation \ref{eqn:vweights} using the expression below:
$$
P_i' = P_i^\gamma
$$
and thus introduce a parameter, $\gamma$, that allows us to smoothly interpolate between random and farthest point sampling \cite{wt-metad,wte}.  In particular, and as shown in the bottom part of figure \ref{fig:staged-algorithm}, when $\gamma < 1$ the procedure is more likely to select landmarks in the vicinity of the densely sampled regions of the space.  By contrast setting $\gamma>1$ encourages the algorithm to ignore the underlying probabilities and to pick a set of landmarks that are more uniformly distributed over the space. 

A function that provides an implementation of this so-called well tempered farthest point sampling algorithm and that takes as input the value of the $\gamma$ parameter, {\it g}, the final number of landmarks required, {\it N}, a list of random vectors, {\it R} and their associated weights, {\it W}, in python is provided below:

\singlespacing
\begin{lstlisting}[caption=Selecting landmarks using the well tempered farthest point sampling algorithm]
def wtfps_landmark_selection( g, N, R, W ) :
    K = int(sqrt(len(R)*N))
    # Select K landmarks using FPS
    fps_l = farthest_point_sampling( K, R, W )
    # Calculate voronoi weights of fps landmarks and assign each of 
    # the input random vectors to its associated voronoi polyhedron
    fps_w, fps_p = voronoi_weights( fps_l, R, W )
    # Modify the weights.  We assume here that the sum of all 
    # the weights in W is equal to one 
    for w in fps_w : w = w**g
    # Create a list containing the indices of the voronoi polyhedra
    fps_i = []
    for i in range(0,k) : fps_i.append(i)
    
    # Now actually select the final landmarks
    landmarks = []
    for i in range(0,N):
        # Get the index of the Voronoi polyhedron from which the 
        # landmark will be selected
        myv = select_random( 1, fps_w, fps_i )
        # Create a list of weights for all the random vectors in 
        # this polyhedron.  All these weights should be set equal 
        # to one. 
        poly_weights = len(fpos_p[myv[0]])*[1]
        # Now select one of the random vectors in the 
        # previously-selected Voronoi polyhedron
        selection = select_random( 1, poly_weights, fps_p[myv[0]] )
        # Add the selected landmark to the final list.  Notice that 
        # code should should be added here so that one random 
        # vector is not added to the list of landmarks multiple 
        # times. 
        landmarks.append( selection[0] )
    
    return landmarks
\end{lstlisting}
\doublespacing

To summarize we often have to run these dimensionality reduction algorithms on a subset of landmark points from the input data set as these algorithms are expensive.  There are three methods we can use to select landmarks:

\begin{enumerate}
    \item {\bf Random sampling} which involves selecting points at random from the input data set.  
    
    \item {\bf Farthest point sampling} which gives us a set of widely spread landmarks.
    
    \item {\bf Well tempered farthese point sampling} which provides a single parameter $\gamma$ that allows us to interpolate between random and farthest point sampling  
\end{enumerate}

In addition, we can ascribe a weight to each of the landmark points we select by using a procedure based on Voronoi diagrams.  This procedure allows one to recover the information on the probability distribution $P(\mathbf{X})$ that is encoded in the distribution of the input random vectors.

\subsection{Step 4: Dimensionality reduction}
\label{sec:dim-red}

In the preceding three sections we have discussed how we can run molecular dynamics or enhanced sampling calculations to generate biomolecular trajectories.  We then discussed how the microscopic states the trajectory samples from can be represented using either a matrix that measures the dissimilarities between each pair of input trajectory frames or by using one high-dimensional vector of structural fingerprints to represent each frame from our trajectory.  Knowing that each trajectory frame can be represented using a high-dimensional vector is critical when it comes to understanding how these dimensionality reduction algorithms work. In fact, many of these algorithms work by orthogonalizing and rotating the basis in which these fingerprint vectors are represented so that the first few vectors in the new basis set describe the majority of the variability in the input data set.  The fact that this mode of operation is true of algorithms such as principal component analysis (PCA), which take the fingerprint vectors as input, is obvious \cite{PCA_book}.  What is less obvious, however, is that methods such as metric multidimensional scaling (MDS), which take a matrix of dissimilarities as input, also work in this way because, as discussed in section \ref{sec:represent}, we can convert any matrix of dissimilarities into a set of high-dimensional, fingerprint vectors \cite{MDS_book}.

\begin{figure}
    \centering
    \includegraphics[width=0.8\textwidth]{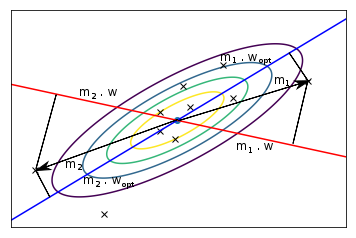}
    \caption{Figure illustrating how the PCA algorithm works.  Each of the black crosses represents one of the $n$ centered fingerprint vectors that are input into the algorithm.  We can calculate the projection of these vectors on any arbitrary vector, $w$.  In the figure we show the projection of two of the fingerprint vectors onto the vector that is indicating using the red line.  The blue line indicates the vector for which the sum of the squares of all these projections is maximized.  The projections of the two fingerprint vectors on this optimal direction are clearly larger than the projections on the red line.  The contour plot in the background of the figure gives a set of isocontours for the function $r^2 = \mathbf{x}^T \mathbf{\Sigma}^{-1} \mathbf{x}$, where $\Sigma$ is a covariance matrix that is calculated from the set of fingerprint vectors.  These isocontours have an elliptical shape and it is clear that the blue line runs parallel to the principal axes of the ellipse.}
    \label{fig:pca-explain}
\end{figure}

Clearly, given the arguments in the previous paragraph, much about dimensionality reduction algorithms can be gleaned from an understanding of the PCA algorithm.  To understand how this algorithm works consider the $n$ centered, fingerprint vectors indicated using the black crosses in figure \ref{fig:pca-explain}.  The coordinates of each of these black crosses can be included in a fingerprint vector that has $m$ components in total.  We can thus put all these vectors into an $n \times m$ matrix, $\mathbf{M}$ that has one fingerprint vector in each of its rows.  We can then calculate the projections of the $n$ fingerprint vectors in $\mathbf{M}$ on any arbitrary $m$-dimensional, unit vector $\mathbf{w}$ using:
\begin{equation}
\mathbf{t} = \mathbf{M} \mathbf{w}
\label{eqn:eq-t}
\end{equation}
This process of taking projections on an arbitrary vector, $\mathbf{w}$ is illustrated in figure \ref{fig:pca-explain} for two of the fingerprints.  The projections of these two fingerprints on the vector, $\mathbf{w}$, which is shown as a red line, are indicated.  When we do the operation above for all of the fingerprint vectors we obtain an $n$-dimensional vector, $\mathbf{t}$, than contains the $n$ projections.  Furthermore, the squared norm of this vector $|\mathbf{t}|^2$ is only large when the unit vector $\mathbf{w}$ encodes a great deal of the variability for the vectors in $\mathbf{M}$.   Performing dimensionality reduction effectively is thus a matter of finding the unit vector $\mathbf{w}$ for which the vector $\mathbf{t}$ is maximal.  In other words, we search over all possible unit vectors, $\mathbf{w}$, and solve the following optimization problem:
$$
\arg \max_{|\mathbf{w}|=1} \left\{ \mathbf{w}^T \mathbf{M}^T \mathbf{M} \mathbf{w} \right\}
$$
In figure \ref{fig:pca-explain} the optimal choice for the vector $\mathbf{w}$ is shown as a blue line.  As you can see the projections of the two chosen points on this blue line are both larger than the projections on the red line.  This optimal choice for the vector $\mathbf{w}$ can be easily found by remembering that the fingerprint vectors in $\mathbf{M}$ are centered and that as such the matrix $\mathbf{M}^T \mathbf{M}$ is nothing more than the $m \times m$ covariance matrix, $\mathbf{C}$.  We can thus reformulate the problem as an optimization of $\mathbf{w}^T\mathbf{C}\mathbf{w}$ subject to the constraint that $\mathbf{w}^T\mathbf{w}=1$ and use the method of Lagrange multipliers.  When employing this method we seek to find stationary points for the following function:
$$
L(\mathbf{w}) = \mathbf{w}^T\mathbf{C}\mathbf{w} - \lambda ( \mathbf{w}^T\mathbf{w}-1 ) 
$$
These stationary points are the vectors, $\mathbf{w}$, that satisfy:
$$
\frac{ \textrm{d}L(\mathbf{w})}{\textrm{d}\mathbf{w}} = \mathbf{C}\mathbf{w} - \lambda \mathbf{w} = 0 \qquad \rightarrow \qquad \mathbf{C}\mathbf{w} = \lambda\mathbf{w}    
$$
What we thus find is that the vector, $\mathbf{w}$, with the largest value for $|\mathbf{t}|^2$ is the eigenvector corresponding to the principal  eigenvalue of the covariance matrix, $\mathbf{C}$.

This process is even simpler when a method such as MDS is performed as we have already seen in section \ref{sec:represent} how we can generate vectors of fingerprints from a $n \times n$ dissimilarity matrix by centering and then diagonalising this matrix.  We could in theory take the $n$ fingerprints that we extract by this procedure and construct an $n \times n$ matrix of data points in this case too, $\mathbf{M}$.  Furthermore, we could then multiply $\mathbf{M}$ by its transpose to obtain a covariance matrix to diagonalize.  Performing these additional steps really is an exercise in futility, however, as the covariance matrix contains the same information as the projections. The projections you would get after applying PCA would thus be identical to the first few rows of the fingerprint vector $\mathbf{V}\Lambda^{\frac{1}{2}}$ that was discussed at the end of section \ref{sec:represent}.

These linear dimensionality-reduction techniques, PCA and MDS, have been part the toolkit data scientists use to analyze data for many years.  It is thus hardly surprising that researchers studying the behavior of biomolecules were quick to apply them to the trajectories that they had extracted \cite{pca-traj-1,pca-traj-2,pca-traj-3}.  The results that were obtained when they performed these analyses, however, were mixed.  One problem was that the first few eigenvectors of the covariance matrix often did not appear to encode the majority of the information about the distribution of the points in the high dimensional space.  In other words, when the principal eigenvector of the covariance matrix was inserted into equation \ref{eqn:eq-t} the norm of the vector $\mathbf{t}$ that emerged was often found to not be very large.  Consequently, much of the information contained in the trajectory was thrown away when the data was projected on the first few eigenvectors of covariance matrix. 

One theoretical justification for using PCA to analyze biomolecular trajectories is a belief that the folded state of a biomolecule is at the bottom of a quasi-harmonic basin in a potential energy landscape.  If this were the case the points visited during the trajectory would be distributed in accordance with a multivariate Gaussian and the PCA eigenvectors would be very similar to those of the Hessian matrix at the minimum in the landscape.  When comparisons were performed between the eigenvectors extracted from a PCA analysis of a trajectory and the eigenvectors extracted from the Hessian matrix of the optimal structure of the protein, however, little similarity between the first few eigenvectors of these matrices were found \cite{pca-problems}.  It was thus concluded that the biomolecules were doing more than simply fluctuating around a single, quasi-harmonic basin in a high-dimensional potential energy landscape.

\begin{figure}
    \centering
    \includegraphics[width=0.8\textwidth]{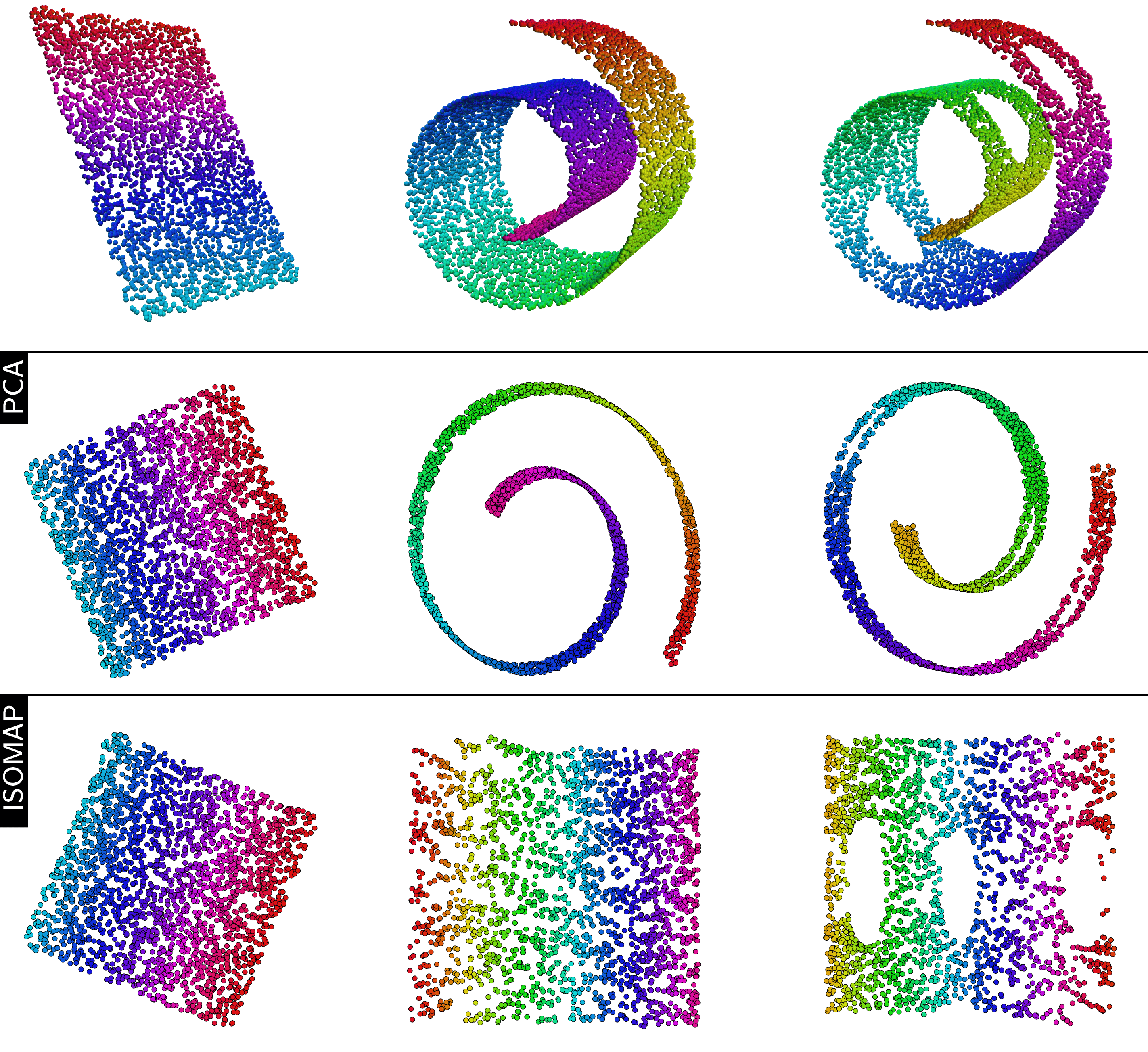}
    \caption{Figure illustrating how PCA and isomap perform on model data.  The top row of the figure illustrates the original data sets.  All three data sets are three dimensional.  In the first data set the model data all lies on a two dimensional plane that is embedded in the three dimensional space.  This structure can thus be found using both PCA and isomap.  In the second data set all the points lie on a non-linear manifold.  As you can see, while isomap is able to unroll this curved manifold and display the relationship between the data points in the plane, PCA is not.  The third data set resembles the second but now there are three circular regions in the curved manifold that are not sampled.  Unsurprisingly, PCA is still unable to produce a projection of this data that recognizes structure of the manifold.  In addition, there are some difficulties with isomap. In particular, the un-sampled regions do not appear to be circular in the projection and are instead elliptical because, as discussed in the text, the presence of the poorly sampled regions ensures that the length of the shortest path through the graph connecting two points is no longer equal to the geodesic distance between those points.  All three data sets were generated by mapping a set of two-dimensional input data points onto the three dimensional manifold of interest.  In the figures above the points are therefore colored according to the values of one of these input coordinates.}
    \label{fig:manifolds}
\end{figure}

An alternative to these linear dimensionality reduction algorithms emerged in the early 2000s with the development of two new algorithms for manifold learning - locally linear embedding \cite{lle} and isomap \cite{isomap}.  The difference between what these methods could do and what can be done with the conventional formulations of PCA and metric MDS is illustrated in figure \ref{fig:manifolds}.  As you can see from the figure the linear methods are able to determine whether the data points all lie on a hyper plane in the high dimensional space.  The non-linear methods, however, are able to determine whether the points lie on a curved manifold - a structure that would not be detected with the linear methods.  In isomap these non-linear structures are found by using the geodesic distances between configurations in place of the euclidean distances that are used in metric MDS.  Consequently, when the resulting matrix of geodesic distances is then centered and diagonalized using the techniques discussed in section \ref{sec:represent}, the fingerprints that emerge give the projections of the structures on the curved space. 

Isomap has been used to analyze trajectory data on biomolecules \cite{isomap-traj-1,isomap-traj-2,isomap-traj-3} but some of the earliest advocates of this approach seem to have now moved on to other algorithms \cite{traj-diffmaps-2,traj-diffmaps-3}.  That there are problems with isomap is well established \cite{donoho1,donoho2,usi-guy}.  Most of these problems arise because of the way the geodesic distances between points are actually computed.  In essence, to calculate the geodesic distance a graph is constructed from the data by connecting two data points if they are within a certain cutoff distance of each other.  The geodesic distance between two points A and B is then found by finding the shortest path through this graph that connects A and B using Dijkstra's algorithm \cite{djkstra} or the Floyd-Warshall algorithm \cite{floyd-warshall}.  The problem with this approach is that, as shown in bottom right panel of figure \ref{fig:manifolds} it works poorly if there are regions of the manifold that are not sampled because the shortest path through the graph, unlike the true geodesic path, has to go around the poorly sampled region.  In addition, and for similar reasons, isomap is also not always effective if there is noise in the directions that are locally orthogonal to the low-dimensional manifold. 

Another non-linear dimensionality that has been used to examine biomolecular trajectories is diffusion maps \cite{diffmap-1,diffmap-2,diffmap-3}.  There have been some promising results \cite{traj-diffmaps-1,traj-diffmaps-4} using this method although some non-trivial modifications are required in order to get this method to work effectively for trajectory data.  In the limited space we have in this chapter we cannot really do justice to the literature on using diffusion maps to analyze trajectory data and would instead direct the interested reader to the following review \cite{clementi-review}.  In the comparisons that follow we have used the related but simpler technique of Laplacian Eigenmaps \cite{laplacian-eigenmaps} in place of diffusion maps.  Much like isomap this algorithm starts by constructing a graph that connects all the data points.  In the simplest version of Laplacian Eigenmaps this is done by constructing a matrix $\mathbf{P}$ which has element $(i,j)$ equal to one if point $i$ and point $j$ are within a certain distance of each other.  In what follows, however, we calculated the $k$ nearest neighbours for each of our data points and set the matrix elements that corresponded to these neighborhood relations to one and all other matrix elements to zero.  We thus introduced a sort of local scale when constructing the graph.  It is worth noting that it is possible to make further modifications to Laplacian Eigenmaps, which make the embedding generated by this algorithm more like that generated by diffusion maps (see Note \ref{note:diffusion-thing}).  To be clear, however, we did not use these particular modifications in what follows.  

In Laplacian Eigenmaps the Laplacian, $\mathbf{L}$ of the weighted graph, $\mathbf{P}$, that is constructed in the first stage is computed using:
$$
\mathbf{L} = \mathbf{D} - \mathbf{P} \qquad \textrm{where} \qquad 
\mathbf{D}_{ij} = 
\begin{cases}
\sum_{j\ne i} P_{ij} & \textrm{if} \quad i=j \\
0 & \textrm{otherwise}
\end{cases}
$$
Once the Laplacian is computed the random-walk-normalized Laplacian is constructed using:
\begin{equation}
\mathbf{L}^{(rw)} = \mathbf{D}^{-1} \mathbf{L}
\label{eqn:laplacian}
\end{equation}
A matrix with low dimensional projections for the $M$ input points in its rows, $\mathbf{X}$, is obtained from this matrix by diagonalizing $\mathbf{L}^{(rw)}$, discarding the lowest eigenvalue and its corresponding eigenvector and by then taking the eigenvectors corresponding to the $N$ lowest eigenvalues that remain, placing them in a $M \times N$ matrix $\mathbf{V}$ and computing:
$$
\mathbf{X} = \mathbf{D} \mathbf{V}
$$

The approach used in diffusion maps is similar to that outlined above for Laplacian Eigenmaps (see Note \ref{note:diffusion-maps}).  Furthermore, the mathematical theory behind both methods is rooted in the theory of discrete time Markov chains.  In particular, these algorithms both assume that the matrix $\mathbf{P}$ can be used to model the rates of diffusion between the input high-dimensional vectors (see Note \ref{note:chapman}).  They then use a combination of the spectral decomposition and the Chapman-Kolmogorov relationship to conclude that diffusion along the eigenvectors whose corresponding eigenvalues are large is slow and that diffusion along the eigenvalues whose corresponding eigenvalues are small is fast.  For diffusion maps constructing projections using the eigenvectors whose corresponding eigenvalues are large therefore ensures that if the modelled rate of diffusion between two points is slow their projections appear far apart.  Furthermore, the same holds for the eigenvectors whose corresponding eigenvalues are small when Laplacian Eigenmaps is used (see Note \ref{note:laplacian}).

The researchers that have used diffusion maps to analyze trajectory data have found that they obtain the best results when they use a locally scaled variant which assumes that diffusion is more rapid in regions of the energy landscape that are sampled more sparsely. In other words, when this locally scaled diffusion maps technique \cite{traj-diffmaps-2,traj-diffmaps-3} is employed it is assumed that diffusion is rapid when the bio-molecule is close to a transition states and slow when it is inside a basin. 
Another algorithm that uses a scale parameter whose value changes based on the local-density of the data is t-distributed stochastic neighbor embedding (t-SNE) \cite{tsne}.  In this method one begins by computing a matrix of conditional probabilities:
\begin{equation}
\mathbf{P}_{j|i} = \frac{\exp\left( - \frac{ | \mathbf{X}_i - \mathbf{X}_j |^2 }{ 2\sigma_i^2 } \right) }{\sum_{k \ne i} \exp\left( - \frac{ | \mathbf{X}_i - \mathbf{X}_k |^2 }{ 2\sigma_i^2 } \right) }  
\label{eqn:tsne}
\end{equation}
The elements of this matrix give a measure of the conditional probability that a data point $\mathbf{X}_i$ would pick a second data point $\mathbf{X}_j$  as its neighbor if neighbors were picked in proportion to their probability density under a Gaussian centered at $\mathbf{X}_i$.  This matrix is not symmetric but a symmetric matrix can be constructed from it using:
$$
\mathbf{P}_{ij} = \frac{\mathbf{P}_{j|i} + \mathbf{P}_{i|j} }{2}
$$
t-SNE then constructs projections, $\mathbf{y}_i$ for each of the input data points by minimizing the Kullback-Leibler divergence between the distribution $\mathbf{P}_{ij}$ and a second distribution:
$$
\mathbf{Q}_{ij} = \frac{( 1 + |\mathbf{y}_i - \mathbf{y}_j|^2)^{-1} }{ \sum_{k\ne j} ( 1 + |\mathbf{y}_i - \mathbf{y}_k|^2)^{-1} }
$$
This distribution is computed from the distances between the projections of the points and the final Kullback-Leibler divergence is computed using:
$$
KL( \mathbf{P} || \mathbf{Q} ) = \sum_{i \ne j} \mathbf{P}_{ij} \log \left( \frac{ \mathbf{P}_{ij} }{ \mathbf{Q}_{ij}} \right) 
$$

As you can see the local scale parameters for the data enter into this procedure through equation \ref{eqn:tsne}.  To calculate these local parameters the user specifies a parameter known as the perplexity, which can be interpreted as a smooth measure of the effective number of neighbors each of the high dimensional data points will have.  Consequently, and much like the scale parameter in the locally scaled version of diffusion maps, the $\sigma$ parameters that appear in equation \ref{eqn:tsne} will be small for those points that are in the densely sampled basins in the energy landscape and large in the transition regions between basins where the sampling is assumed to be much more sparse. 

\begin{figure}
    \centering
    \includegraphics[width=\textwidth]{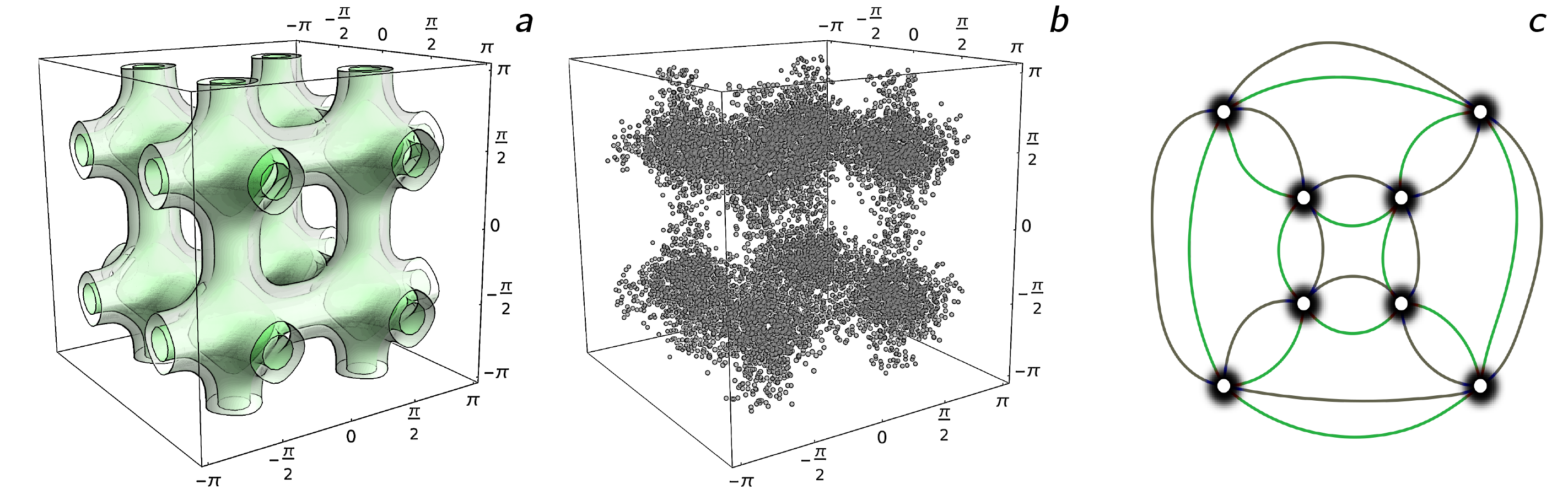}
    \caption{Figure illustrating the form of the data set that was used in the development of the sketch-map algorithm.  The left-most panel of the figure shows the probability distribution from which points were sampled, while the central panel shows the points that were sampled from the distributions and analyzed using the various dimensional reduction algorithms.  The right most panel shows an idealized projection of the data in two dimensions.  As you can see from the left panel the distribution has eight modes and the distribution is periodic in all three directions.  Consequently, each pair of modes is connected by two distinct pathways.  It is this structure that we would thus like to see in the low-dimensional projection.}
    \label{fig:3dbox-data}
\end{figure}

The final dimensionality reduction technique that we will discuss is our own method sketch-map \cite{sketch-map,fieldcvs,smap-robust}.  Furthermore, this technique will be the subject of much of the rest of this chapter.  We developed sketch-map while studying the data from a simulation \cite{recon} of a twelve-residue sequence of alanines \cite{wales-ala12} because when we tried to use the algorithms described in the previous paragraphs to project these trajectories we got a low dimensional projection that was not particularly informative.  In particular, we never observed a wide gap between the norms of the $\mathbf{t}$-vectors that were obtained when any two neighboring eigenvectors, $\mathbf{w}$, were inserted into equation \ref{eqn:eq-t}.  Instead we observed a steady decline in the values of the norms of the $\mathbf{t}$-vectors for the various eigenvectors and thus concluded that the information in this data set was spread out over all over the high dimensional space and that as such the conventional techniques would not work.  We thus sought to develop a three dimensional data set, which we knew we could not project using any of the algorithms outlined above in the hope that if we were able to develop an algorithm that could give us a meaningful projection of this data it would also give us meaningful information on our ala12 trajectories.  The data set we developed for this purpose is shown in central panel of figure \ref{fig:3dbox-data}.  This data was generated by randomly sampling points from the probability distribution:
$$
p(x,y,z) = \exp\left( 3[3-\sin^4(x) - \sin^4(y) - \sin^4(z)] - 1\right)
$$
An isosurface in this probability density is shown in the left panel of figure \ref{fig:3dbox-data}.  What makes data generated from this distribution so difficult to project is the topology of this probability distribution.  The energy landscape that underpins this probability distribution has eight basins and most of the points that are generated are samples from these basins.  Each pair of basins is then connected by two transition pathways, one which runs through the center of the box and one which runs through the periodic boundary.  An ideal two dimensional projection of this data would thus look something like the cartoon shown in the right-most panel of figure \ref{fig:3dbox-data}.  

\begin{figure}
    \centering
    \includegraphics[width=\textwidth]{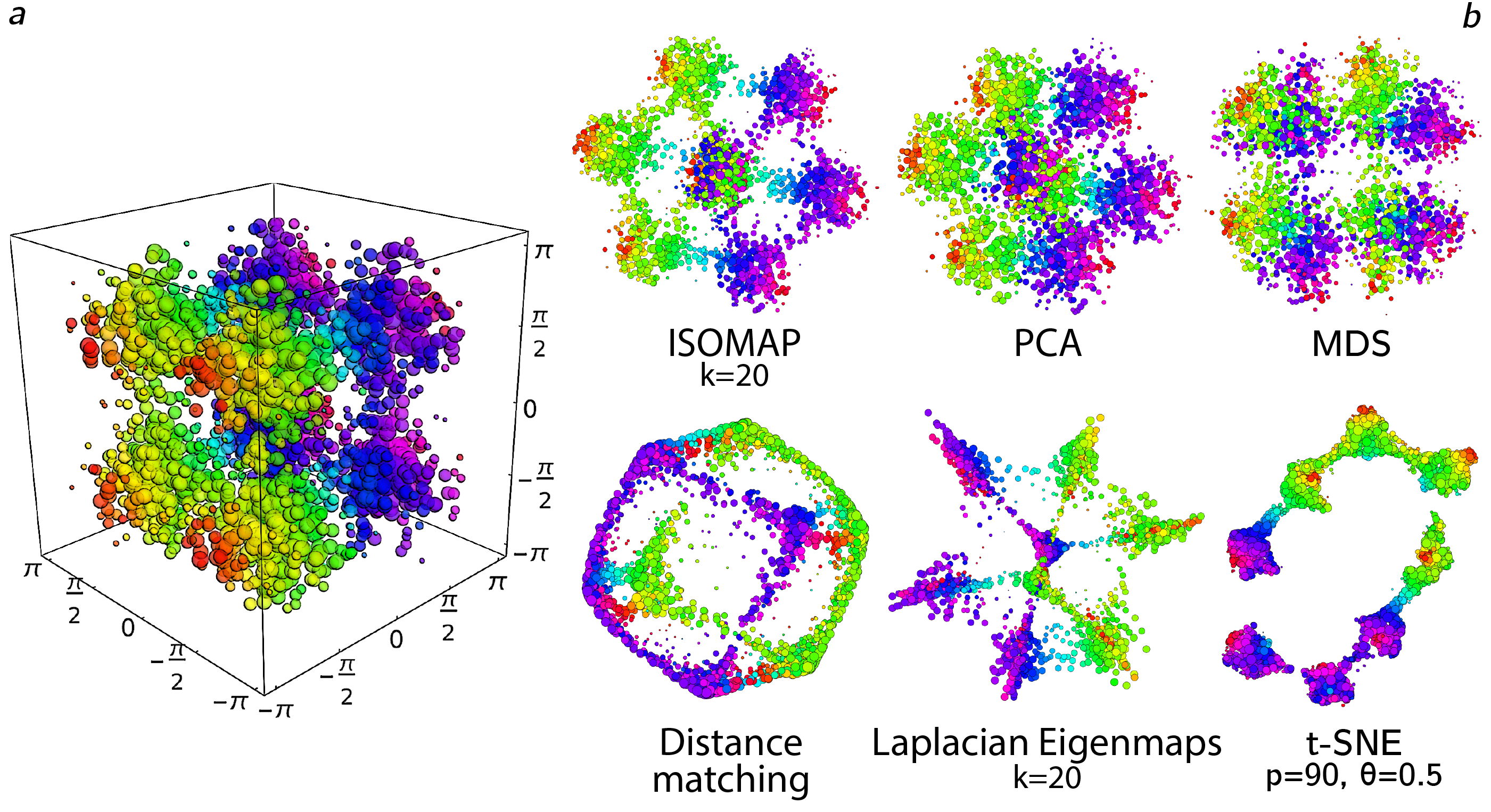}
    \caption{Figure showing the projections of the data set that was introduced in figure \ref{fig:3dbox-data} using the dimensionality reduction algorithms that have been discussed in the text.  The left panel shows the three dimensional data set that was projected once more but the points are now colored in accordance with the value of one of the three high-dimensional coordinates.  The points in each of the projections are colored in the same way.  Notice that none of these projections are similar to the ideal projection shown in the right panel of figure \ref{fig:3dbox-data}.  In particular, none of the projections allow one to determine that each basin in the landscape is connected by two transition pathways.  Hyperparameters for each of the algorithms are given above.}
    \label{fig:3dbox-actual-proj}
\end{figure}

Projections of the data set in figure \ref{fig:3dbox-data} were constructed using the implementations of the algorithms described in the previous paragraphs that are in SciKit Learn \cite{scikit-learn}.  For isomap and Laplacian Eigenmaps we constructed a graph connecting all the points using a $k$-nearest neighbor approach with $k=20$.  For t-SNE we used a perplexity value of 90 and the Barnes-Hut implementation in SciKit learn with an angular size of 0.5.  The final results are shown in figure \ref{fig:3dbox-actual-proj}.  As you can see the performance of all of these algorithms is far from satisfactory.

Distance matching is arguably the best performing of the algorithms tested in figure \ref{fig:3dbox-actual-proj} as in the projection generated using this method it is clear that at least some of the basins are connected by two different pathways.  When this algorithm is used all the linear algebra discussed in section \ref{sec:represent} is discarded and projections, $\mathbf{x}$, are found by minimizing the stress function:
\begin{equation}
\chi^2(\mathbf{x}) = \sum_{i \ne j} w_i w_j (D_{ij} - d_{ij})^2 
\label{eqn:dmatch}
\end{equation}
using an interactive algorithm such as steepest descent or conjugate gradients.  In equation~\ref{eqn:dmatch} $D_{ij}$ is the dissimilarity between the high-dimensional vectors of fingerprints for configurations $i$ and $j$ and $d_{ij}$ represents the distance between the corresponding projections of these two points.  A further advantage of this algorithm over those discussed thus far is that the weights discussed in sections \ref{sec:collect} and \ref{sec:landmarks} can be included in the stress function.  For these reasons we thus chose this method as our start point when developing sketch-map.

There is a rich literature on generating low-dimensional projections of high dimensional data by optimizing stress functions such as the one in equation \ref{eqn:dmatch} \cite{MDS_book,usi-guy}.  Many of these algorithms work by giving each distance that appears in the stress function a weight, $w_{ij}$.  By adjusting the weights of these distances one can then force the algorithm to focus its attention on getting the distances between particular pairs of projections to match the dissimilarities between a particularly important pair of high-dimensional fingerprints.  Alternatively, a second class of algorithm focuses on ensuring that the distances between the projections gives information on the ordering of the dissimilarities between the high-dimensional vector of fingerprints \cite{MDS_book}.  We mention these algorithms here not because we need to focus on their details but rather because of what they tell us about how this business of dimensionality reduction has been approached in other fields.  In short, researchers have used their intuition about the data being studied to adjust the stress function that is optimized by the algorithm in a way that downplays the uninteresting information contained in the high-dimensional distribution.  By doing so they have thus developed algorithms that focus on reproducing, in the low dimensional projection, the information from the high-dimensional data set that they believe is important based on their intuition about the problem.  This realization is important in the context of sketch-map as this algorithm does not produce an isometric mapping of the high dimensional space as is done in many other dimensionality reduction algorithms.  Instead, much of the information on the disposition of the points in the high-dimensional space is discarded so that the algorithm can focus on producing a low-dimensional projection that contains the most pertinent information.

In practice a sketch-map projection, $\mathbf{x}$, is generated by optimizing the following stress function:
\begin{equation}
\chi(\mathbf{x}) = \sum_{i \ne j } w_i w_j [ F(D_{ij}) - f(d_{ij}) ]^2
\label{eqn:smap-stress}
\end{equation}
As in equation \ref{eqn:dmatch} $D_{ij}$ here is the dissimilarity between the high-dimensional fingerprints for configurations $i$ and $j$ respectively and $d_{ij}$ is the distance between the projections of these two points.  At variance with equation \ref{eqn:dmatch}, however, these two distances are transformed by two sigmoid functions of the form:
\begin{equation}
F(x) = 1 - (1 + (2^{a/b} - 1)(r/\sigma)^a)^{-b/a}
\label{eqn:smap-sigmoid}
\end{equation}
which have the same value for the $\sigma$ parameter but different values for the $a$ and $b$ parameters.  These two functions have a value that is close to zero for values of $x$ that are much less that $\sigma$ and a value that is close to one for values of $x$ that are much greater than $\sigma$.  Incorporating these two function in the stress function in equation \ref{eqn:smap-stress} ensures that the algorithm focuses most of its attention on reproducing the dissimilarities that are close to $\sigma$ when constructing projections.  
Meanwhile, if points are separated by less than $\sigma$ in the high dimensional space their sketch-map projections will appear very close together.  In addition, the projections of points that are very far apart in the high-dimensional space can be almost arbitrarily far apart.  In other words, sketch-map focuses on reproducing proximity information from the high-dimensional data set.  It ensures that points that are closer than a characteristic distance are mapped close together, while simultaneously ensuring that the farther apart points are well separated in the projection. 

\begin{figure}
    \centering
    \includegraphics[width=0.8\textwidth]{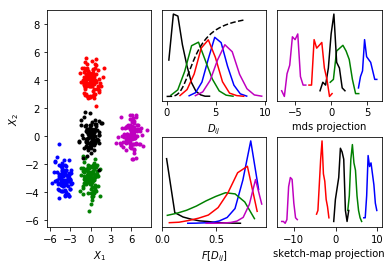}
    \caption{Figure illustrating the purpose of the sigmoid functions in sketch-map.  The right most panels of this figure show the 1D-projections of the model data in the left most panel of the figure that are generated by MDS (upper) and sketch-map (lower).  The 2D model data in the left most panel was generated by sampling points from five normal distributions.  The points in the left panel are colored according to the distribution they were sampled from.  Furthermore, to illustrate the projections of each of the basins in the 1D space we show the histograms for the projections of the points in each of the five basins.  The upper central panel shows the distribution of distances between the points that are shown in black in the left panel and each of the other sets of points in the panel.  In this panel we also show the sigmoid function that was used in sketch-map to transform these distances using a dashed line.  The lower central panel shows the histograms for the transformed distances.  It is clear from these histograms of transformed distances that the sigmoid function squeezes together points that belong to the same feature while spreading out points that belong to different features.      }
    \label{fig:smap-operation}
\end{figure}

The reason sketch-map focuses on reproducing the dissimilarities that have values that are close to $\sigma$ when constructing the projection is that these dissimilarities are considered to be the most important in terms of understanding the structure of configuration space.  It is assumed that the parts of configuration space that are sampled in any trajectory are clustered around energetic basins.  These basins are then connected by a spiders web of transition pathways.   What one would thus like to do with the dimensionality reduction algorithm is to visualize the connections between the energetic basins.  The internal structure of the basins, which is less interesting, should be collapsed in the projection and any points that are in basins that are very far apart should be projected very far apart so that it is clear to see these basins are not connected by a transition pathway.  The degree to which sketch-map succeeds in this regard is illustrated in figure \ref{fig:smap-operation}.  To generate this figure we generated the data shown in the right most panel by sampling a series of points from 5 normal distributions.  These normal distributions were arranged in the two dimensional space so that it would be difficult to produce a one-dimensional projection of the data using MDS.  Furthermore, the points in the left panel of the figure are colored in accordance with the Gaussian they were sampled from.  As you can see from the top right panel of the figure the projection we get using MDS is not so revealing.  To generate this panel we took the projections of each of the data points in each of the basins a generated a separate histogram for each of the basins using kernel density estimation.  This procedure gave us a sense of the shape of each of the projected basins and as you can see there are substantial overlaps between the various basins when projections are constructed using MDS.  These overlaps are not present for the sketch-map projections that are shown in the lower right panel, however.  The reason sketch-map performs better is illustrated in the two central panels.  The upper figure here shows the distribution of the distances between the points that are shown in black in the left panel of the figure and each of the other sets of points in the figure.  There is considerable overlap between the green red and black distributions, which is why in the MDS projections the black histogram overlaps with the green and red histograms.  The upper panel in figure \ref{fig:smap-operation} also shows a dashed line that indicates the sigmoid function (equation \ref{eqn:smap-sigmoid}) that has been used within the sketch-map algorithm.  The lower central panel then shows the histograms for the transformed distances between the points that are shown in black in the left panel of the figure and each of the other sets of points in the figure.  As is clear from the figure the sigmoid converts the majority of the in-basin distances that connect black points to black points to values that are close to one.  Similarly the majority of the distances that connect black points to blue or purple points are converted to one by the sigmoid.  As a consequence during the fitting process sketch-map works hard to ensure that the distances between the black and red and the black and green points are reproduced in the projection.  The black points, meanwhile, are projected closer together than they are in actuality, while the distances between the black and blue and black and purple points are extended in the projection.  The fact that these distances can be distorted in this way is what ensures that each of the basins appear as separate, non-overlapping features in the projection in the lower right panel of figure \ref{fig:smap-operation}.

\begin{figure}
    \centering
    \includegraphics[width=0.8\textwidth]{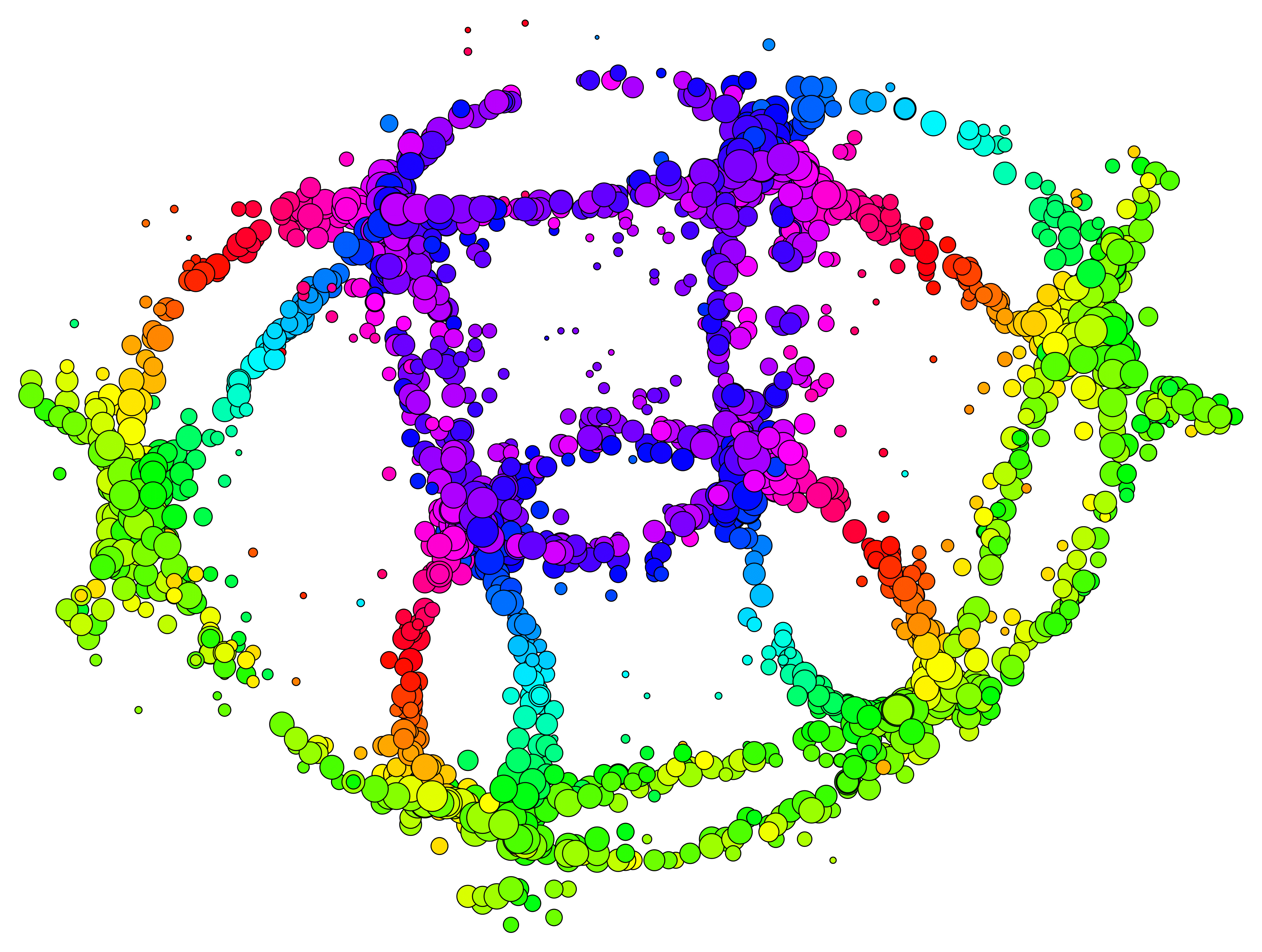}
    \caption{Figure illustrating the projection that is generated by sketch-map of the data set that was introduced in figure \ref{fig:3dbox-data}.  This representation is much closer to the ideal projection that was shown in the right panel of figure \ref{fig:3dbox-data} than any of the representations that were shown in figure \ref{fig:3dbox-actual-proj}.  In particular the two pathways connecting each of the basins are much clearer in the sketch-map representation that is shown above. The hyperparameters used in construcing this projection were $\sigma=2$, $A=2$, $B=10$, $a=2$ and $b=3$.}
    \label{fig:3dbox-smap}
\end{figure}

Figure \ref{fig:3dbox-smap} shows a sketch-map projection of the data from figure \ref{fig:3dbox-data}.  It is clear that the sigmoid functions once again contract each of the basins and thus ensure that the algorithm focuses on reproducing the distances between the various basins.  As a consequence it is much easier to see that there are multiple transition pathways between each pair of basins in the energy landscape.  Admittedly, the projection is still not the ideal configuration shown in the right-most panel of figure \ref{fig:3dbox-data} but it is certainly more revealing than the projections of this data that are shown in figure \ref{fig:3dbox-actual-proj}. 

In this section a lot of detail about the various dimensionality reduction algorithms have been used to analyze biochemical trajectories has been provided.  It is impossible to summarize all this information in a single paragraph but it is worth emphasizing that the differences between algorithms that have been discussed are in the ways that the dissimilarities between the vectors of fingerprints for each configuration are calculated and employed.  Progress has been made and better algorithms have been developed by either:

\begin{enumerate}
    \item Thinking of ways to calculate physically meaningful dissimilarities between configurations.  For example the model of diffusion that is used in diffusion maps notionally ensures that the projection coordinates are the directions along which diffusion is slow.
    
    \item Pragmatically discarding dissimilarities that are thought to be uninteresting when constructing projections as is done in sketch-map.
\end{enumerate}

In other words, the algorithms that work well are those that endeavor to use the known physics of the problem when constructing projections.

\section{Examples}

The previous sections of this chapter have introduced the theory behind a number of dimensionality reduction algorithms.  In the following three sections we will show how these methods have been applied in practice.  We will begin by projecting some data from a simulation of the C-terminal fragment of the immunoglobulin binding domain B1 of protein G of Streptococcus using some of the algorithms that were discussed in the previous section in order to compare their performances.  We will then give a brief survey of the ways in which the sketch-map algorithm has been used by the community.  Finally, we will finish by discussing the challenge of accurate sampling and how sketch-map has been used to enhance sampling.

\subsection{Performance}

\begin{figure}
    \centering
    \includegraphics[width=\textwidth]{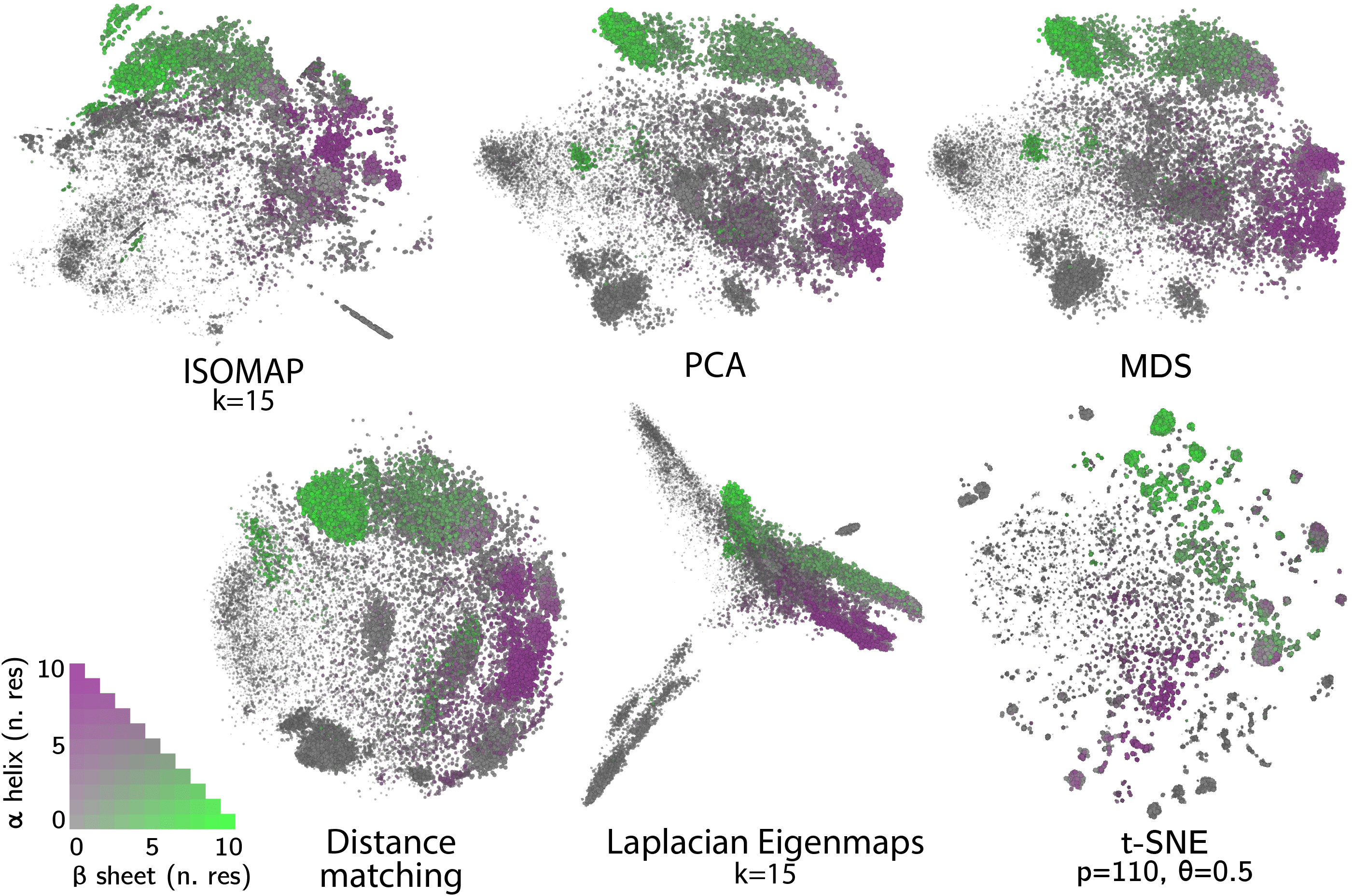}
    \caption{Projections of a parallel tempering trajectory of the C-terminal fragment of the immunoglobulin binding domain B1 of protein G of Streptococcus.  Each of the figures above is a projection of 25311 randomly selected frames from the trajectory of the wild type protein that was calculated in the paper by Ardevol \emph{et al.} \cite{albert-beta1}.  We used the STRIDE algorithm \cite{stride} to determine how many residues had a configuration similar to a beta sheet and how many residues had a configuration similar to an alpha helix for each of the configurations in the trajectory.  In the projections above we have thus coloured the points in each of the projections in accordance with the secondary structure that was observed in the corresponding trajectory frame.}
    \label{fig:smap-comparison}
\end{figure}

In section \ref{sec:dim-red} we showed how the various different dimensionality reduction algorithms that we have discussed fare when projecting some model data.  This was, arguably, not a particularly fair test as the model data was deliberately designed so that sketch-map would outperform the others.  In preparing this section we have thus taken some data \cite{albert-beta1} from a parallel tempering trajectory of the C-terminal fragment of the immunoglobulin binding domain B1 of protein G of Streptococcus and projected it using the various algorithms that were discussed in the previous section. The final results are shown in figures \ref{fig:smap-comparison} and \ref{fig:free-energy-with-snapshots}.

To construct the projections shown in figure \ref{fig:smap-comparison} we took 25311 randomly-selected points from the wild type trajectories that were presented in the paper by Ardevol \emph{et al.} \cite{albert-beta1}.  For each of these configurations we computed the full set of 16 torsional backbone dihedral angles.  Two dimensional projections for each of these 32-dimensional vectors were then generated using the implementations of the various algorithms described in the figure that are available in SciKit Learn \cite{scikit-learn}.  The hyper parameters that we used for each of these algorithms are given in the figure. 

\begin{figure}
    \centering
    \includegraphics[width=\textwidth]{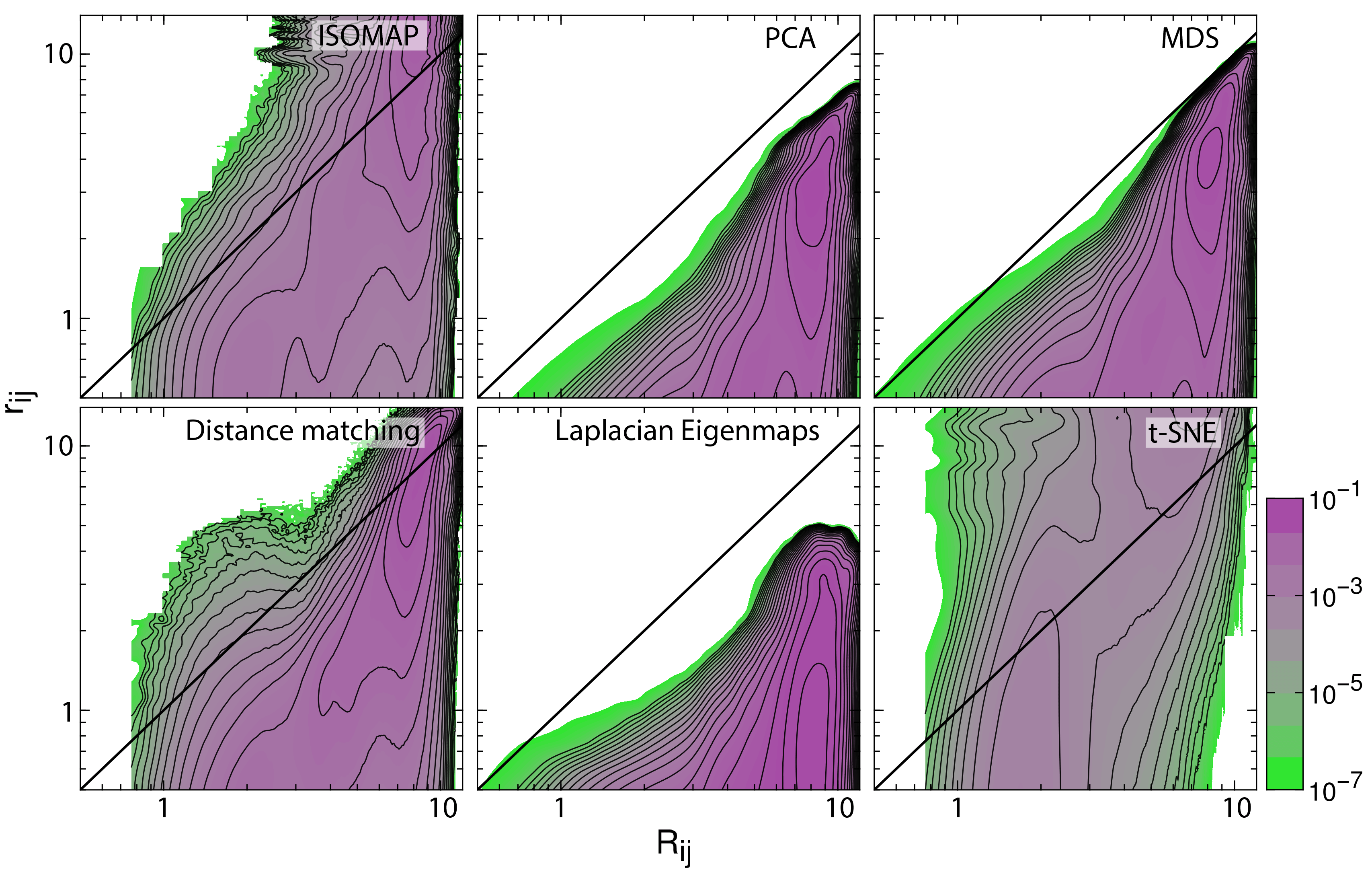}
    \caption{Histograms illustrating the joint probability density function for the dissimilarities between the configurations in the trajectory and the distances between the corresponding projections of these trajectory frames.  The particular projections that have been analyzed here are those that are shown in figure \ref{fig:smap-comparison}.  The black line in each of these figures is the line $R_{ij}=r_{ij}$.  For an ideal projection all the density in these histograms would lie on this line. }
    \label{fig:smap-dist-distribu}
\end{figure}

Before projecting the trajectory we used the STRIDE algorithm \cite{stride} to determine the secondary structure content in each of the frames that was analyzed.  In particular, we counted the number of residues that had a structure that was similar to an alpha helix and the number of residues that had a structure that was similar to a beta sheet.  When constructing the projections in figure \ref{fig:smap-comparison} we thus colored the projections according to the number of residues in the corresponding trajectory frames that appeared to be in a alpha helix configuration and the number of residues that appeared to be in a configuration that resembled a beta hairpin.  Coloring the projections in this way gives us a qualitative way to compare how well each of the algorithms does when it comes to projecting the trajectory data.  What we see is that all the algorithms do a reasonable job of separating the configurations that are predominantly alpha helix like from those that have a structure that is predominantly composed of beta sheets.  In this sense at least then the algorithms all give a reasonable projection of the high-dimensional data.

In section \ref{sec:dim-red} we discussed how the classical MDS and PCA algorithms that were used to construct the top right and top centre panels of figure \ref{fig:smap-comparison} are identical.  The fact that these two projections of the beta hairpin data are very similar is thus perhaps unsurprising.  It is important to note, however, that this similarity persists here even though slightly different representations of the input data were used when constructing these two projections.  In particular, when constructing the MDS projection the input, high-dimensional vectors contained the 32 backbone dihedral angles and distances between these vectors were computed in a way that took the periodicity of these quantities into account.  To run PCA, however, we needed to use 64-dimensional input vectors containing the sines and cosines \cite{dihedral-pca-1,dihedral-pca-2} of the backbone dihedral angles as this algorithm will not work if any of the high-dimensional input variables are periodic.

Although the projections that have been generated using PCA and MDS separate the configurations that resemble alpha helices from those that resemble beta sheets it is clear from figure \ref{fig:smap-comparison} that these projections do not provide an optimal reflection of the distances between the high-dimensional data points.  In section \ref{sec:dim-red} we discussed how these two algorithms find the low-dimensional representation by projecting the data on a two-dimensional plane that is embedded in the high dimensional space.  It is clear from figure \ref{fig:smap-comparison}, however, that many of the high-dimensional points do not lie within this plane as the projection shown in the bottom left of the figure that was generated using the distance matching algorithm is radically different from the PCA and MDS projection.  In particular, the points in this projection are spread out more uniformly across the low dimensional space and some of the clusters that were apparent in the PCA and MDS projections have disappeared.  It is thus clear from these three projections that the trajectory data does not simply lie on two dimensional linear manifold.

Further evidence that the points do not lie on a two dimensional linear manifold is provided by figure \ref{fig:smap-dist-distribu}.  To construct the panels shown in this figure we generated two dimensional histograms and thus estimated the joint probability density function for the dissimilarities between the trajectory frames and the distances between the projections of these configurations.  Furthermore, we constructed these histograms for all of the projections that are shown in figure \ref{fig:smap-comparison}.  The results from PCA and MDS are shown in the middle top and right top panel of figure \ref{fig:smap-dist-distribu} respectively.  For both of these algorithms the distances between the projections of the points are systematically shorter than the dissimilarities between the actual trajectory frames.  The reason these distances are shorter is that for both of these algorithms the distance between any pair of projections is equal to the length of a projection of the vector connecting the two configurations in a two dimensional space.  The lengths of the projections of the vectors connecting the configurations are shorter than the lengths of the original, un-projected and high-dimensional vectors because during the projection operation some components of these vectors are discarded.  Notice that a different behavior is observed when distance matching is used in place of a these linear techniques.  When the projections are found by minimising a stress function using an iterative algorithm the projections the algorithm finds no longer have to lie on a low dimensional linear manifold.  Instead the distance matching algorithm must simply seek to match as many distances to dissimilarities as possible.  In the histogram shown in the bottom left panel of figure \ref{fig:smap-dist-distribu} we thus see that the number of distances between pairs of projections that are larger than the corresponding dissimilarities is roughly equal to the number of distances between pair of projections that are shorter than the corresponding dissimilarities.  Furthermore, the average value for the distances between the projections is approximately equal to the average value for the average dissimilarity.   

Figure \ref{fig:smap-comparison} shows that none of the non-linear dimensionality reduction algorithms that were described in the previous section do much better than the linear methods when it comes to projecting the trajectory data.  In fact the ISOMAP projection that is shown in the top left of the figure bears some similarity with the projections that were generated using PCA and MDS.  The similarity between these two projections suggests that the geodesic distances are similar to the euclidean distances and that the trajectory does not uniformly sample a non-linear manifold in the high dimensional space.  The histogram in the top left hand corner of figure \ref{fig:smap-dist-distribu} suggests that there are differences between the geodesic and the euclidean distances, however.  The figure shows that the distances between the projections of many of the most dissimilar configurations are considerably larger than the dissimilarities between the trajectory frames.  It would seem, therefore, that replacing the euclidean distances with geodesic distances has made a substantial difference but that it is difficult to see this difference just by looking at the projection shown in figure \ref{fig:smap-comparison}.


The projection that was generated using Laplacian Eigenmaps is shown in the bottom center of figure \ref{fig:smap-comparison}. 
The Laplacian Eigenmaps projection has the configurations that resemble alpha helices projected closer to the projections of configurations that resemble the beta sheets than the other projections.  The model of diffusion that underpins this method thus suggests that diffusion between these configurations is relatively rapid.  This makes physical sense as one would expect the slowest process in the system to be diffusion between the folded states and the unfolded states that are projected in the periphery of the map.  If one wishes to examine the relative free energies of the various different folded states, however, this representation may not be optimal.  

It is perhaps not fair to compare the distances between the projections of the points with the dissimilarities for this algorithm as the Laplacian Eigenmaps makes no effort to generate a projection that reproduces these quantities.  The joint probability distribution for the dissimilarities between the trajectory frames and the distances between the corresponding projections that is obtained using this algorithm is nevertheless shown in the bottom middle panel of figure \ref{fig:smap-comparison}.  It is clear that many configurations are projected much closer together than they are in actuality and that the distances between the projection of any two configurations is likely to be close to zero even if the dissimilarity between the two configurations is substantial.  The reason that there are such big mismatches between the distances and the dissimilarities is that when we construct the graph that is used to model the diffusion between the high-dimensional data points each point is connected to its $k$ nearest neighbors.  Two neighboring points can be very far apart, however, particularly in regions of configuration space that are sampled sparsely.  In other words, when using the the Laplacian maps algorithm in the way we have applied it one assumes that the non-linear manifold whose structure one is endeavoring discover using the dimensionality reduction algorithm is sampled relatively uniformly.  This is clearly not true in our case as we know that an MD simulation will sample extensively from the basins in the energy landscape and that the transition states will be weakly sampled.  This uneven sampling of phase space is in fact one reason why the modifications that introduce local scaling parameters into to diffusion maps and that were discussed in section \ref{sec:dim-red} are required when analyzing trajectory data \cite{clementi-review}.    

Another algorithm that introduces a kind of local scaling is t-SNE.  The t-SNE projection of the trajectory data is shown in the bottom right of figure \ref{fig:smap-comparison}. 
This representation is composed of a large number of disjoint clusters and consequently if the free energy surface were projected as a function of these coordinates it would appear very rough.  If one looks more closely, however, the structures in many of these clusters are very similar.  In the representation shown in figure \ref{fig:smap-comparison} for example the configurations that resemble an alpha helix appear to have been split between a number of different basins, which is a very different behavior to that observed for the other representations of the trajectory.  The reason the projection appears this way is clear from the histogram that is shown in the bottom right panel of figure \ref{fig:smap-dist-distribu}.  It would seem that the distances between the projections that are constructed using the t-SNE algorithms are much larger than the dissimilarities between the corresponding trajectory frames.    

A projection of the $\beta$-hairpin trajectory that was generated using sketch-map is shown in figure \ref{fig:free-energy-with-snapshots}.  This projection resembles the projection that was generated using t-SNE in that many clusters in the data have been identified.  At variance with t-SNE, however, all the configurations that resemble alpha helices have been projected in one cluster close to the center of the map, while all the configurations that resemble beta sheets have been projected in a second, different cluster at the center of the map.  High-energy configurations that resemble neither of these two secondary structure types have meanwhile been projected in the periphery of the map.  In other words, for this particular data set sketch-map appears to have generated a projection that has an appearance that is intermediate between that generated by t-SNE and those generated by the other algorithms.   Furthermore, it has done so using a single scale parameter for all points without the need to resort to any form of local scaling.  The reason the sketch-map projection appears this way is clear from the histogram that is shown in the inset in figure \ref{fig:free-energy-with-snapshots}.  This histogram, much like those shown in figure \ref{fig:smap-dist-distribu}, shows the joint probability density function for the dissimilarities between trajectory frames and the distances between their corresponding projections.  The histogram that is observed for sketch-map is similar to the histogram that was observed for t-SNE in that the points that are close together are projected much closer together than they are in actuality.  The distances between the projections of the configurations that are far apart, however, can be much larger than corresponding dissimilarities.  Even so there is a substantial difference between the histograms that are observed with t-SNE and sketch-map.  For sketch-map there is a region around $\sigma=6$ where the majority of the dissimilarities and the distances are very similar.  This behavior occurs because, as discussed in section \ref{sec:dim-red}, the two sigmoid function in the stress function that is optimized within sketch-map ensure that the projection will reproduce the distances in this particular range.  This ability to control the shape of this histogram and by extension the distances that will be reproduced in the projection is the real strength of the sketch-map algorithm.  Sketch-map unlike the other algorithms that have been discussed in this section allows you to pragmatically chose the distances that you would like to accurately reproduce when you construct projections.  Figure \ref{fig:smap-dist-distribu} and the discussions above show that when the other algorithms that have been described in this section are used in place of sketch-map the user has much less control over the distances that are accurately reproduced.          

\subsection{Applications}

In the previous section we discussed the efficacy of the various dimensionality reduction algorithms in terms of whether they could distinguish configurations containing alpha helices from those containing beta sheets.  Given this it is perhaps not unreasonable to ask what purpose is served by using these dimensionality reduction algorithms?  The previous section suggests that we would be better off using CVs that measure the numbers of alpha helices and beta sheets in the protein when analyzing the trajectory that was the subject of the previous section \cite{alpharmsd}.  We would then have a projection of the trajectory that we understand and that therefore is perhaps more physically revealing.

\begin{figure}
    \centering
    \includegraphics[width=\textwidth]{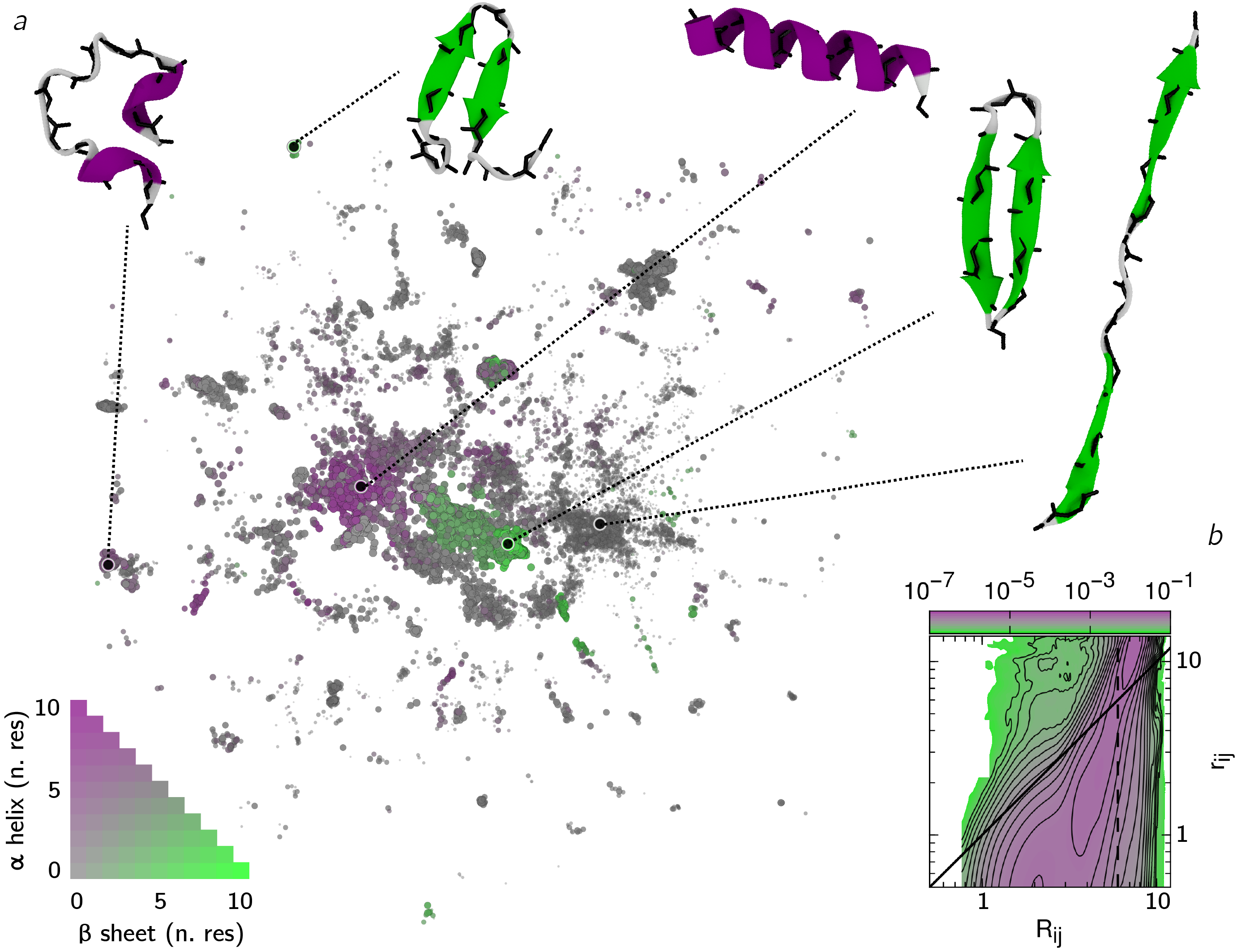}
    \caption{A sketch-map projection for a parallel tempering trajectory of the C-terminal fragment of the immunoglobulin binding domain B1 of protein G of Streptococcus.  The data that was used to construct this projection was taken from the work of Ardevol \emph{et al.} \cite{albert-beta1}.  In particular, the simulations of the wild-type protein.  The initial sketch-map projection here was constructed from 1000 landmark point which were selected using the well tempered farthest point sampling algorithm that was described in section \ref{sec:landmarks} and a gamma parameter of 0.1.  Weights for each of these landmarks were generated using a Voronoi procedure and the sketch-map stress function with parameters $\sigma=6$, $A=8$, $B=8$, $a=2$ and $b=8$ was then optimized to find the landmarks.  Once projections for these landmarks had been found the remainder of the trajectory was projected using the out of sample procedure.  The location at which a number of representative structures are projected has been indicated in the figure.  In addition, we used STRIDE \cite{stride} to determine the number of residues in each configuration that was visited in the trajectory that had the atoms arranged similarly to the arrangement of the atoms in a beta sheet or alpha helix.  As was the case in figure \ref{fig:smap-comparison} the points in the above representation are colored according to the secondary structure that is observed in the corresponding configuration.  In addition, a histogram similar to those in figure \ref{fig:smap-dist-distribu} that shows the joint probability distribution for the dissimilarities between the high dimensional configurations and the distances between the corresponding projections of these points is shown in the bottom right corner of the figure.}
    \label{fig:free-energy-with-snapshots}
\end{figure}

There is certainly some merit to the argument outlined in the previous paragraph.  If you have some clearly defined physical/chemical question to answer then you should display the free energy surface as a function of some CVs that allow you to answer the question you seek to answer.  For example, if you are interested in the relative free energies of the folded and unfolded states of a protein and if you know the structure of the folded state, it is probably best to display the free energy as a function of a CV, such as RMSD, that is small when the structure is folded and that is large when it is not.  After all, and as we have said many times in this chapter, these dimensionality reduction algorithms should not be used to replace your chemical/physical intuitions about the problem.  The problem with chemical intuition, however, is that there are many physical systems for which our intuition is severely lacking \cite{signal,homo-1,homo-2,homo-3}.  For example, there are many so-called intrinsically-disordered proteins that do not have a clear folded state \cite{idp}.  It is thus when studying these types of problems that the insights that can be obtained by performing an analysis using a dimensionality reduction algorithm can prove invaluable.  Dimensionality reduction allows one to extract a visual representation of the ensemble of configurations that have been sampled during the simulation.  The free energy can be projected as a function of these low dimensional coordinates and, because there is a one to one mapping between configurations in the trajectory and the projections of the low dimensional points, you can get some insight into the structures in the various basins that are found in this energy landscape.  An example, where sketch-map has been used to generate this sort of representation is shown in figure \ref{fig:free-energy-with-snapshots} \cite{albert-beta1}.  Notice that we surround the free energy surface with snapshots from the trajectories in this figure and indicate where each of these structures are projected in the low dimensional representation.  This step of working out what structures are projected in each part of the landscape is critical for interpreting the the free energy surfaces when they are output in terms of these types of automated coordinates.  

These automated approaches for generating collective variables show real promise when it comes to investigating how a small perturbation in the conditions can affect the free energy landscape and hence the properties of the system under investigation.  Obviously, any change in the conditions causes the system's Hamiltonian to change.  Even if the change to the Hamiltonian is relatively small, however, there can be a substantial difference in the free energy surface and hence the properties of the perturbed system.  Furthermore, the complicated relationship between the Hamiltonian and the free energy surface makes predicting what changes there will be almost impossible.  These difficulties thus clearly make determining what collective variable to use when visualizing these free energy surfaces extremely challenging.  
By using a dimensionality reduction algorithm to extract a representation from the trajectories, however, you essentially sidestep these problems.  Furthermore, because these algorithms give you an unbiased view of the ensemble of configurations that were sampled during the trajectory, the differences between the perturbed and unperturbed free energy landscapes provides information on changes in the properties of the system that you might not have predicted otherwise. 

\begin{figure}
    \centering
    \includegraphics[width=0.8\textwidth]{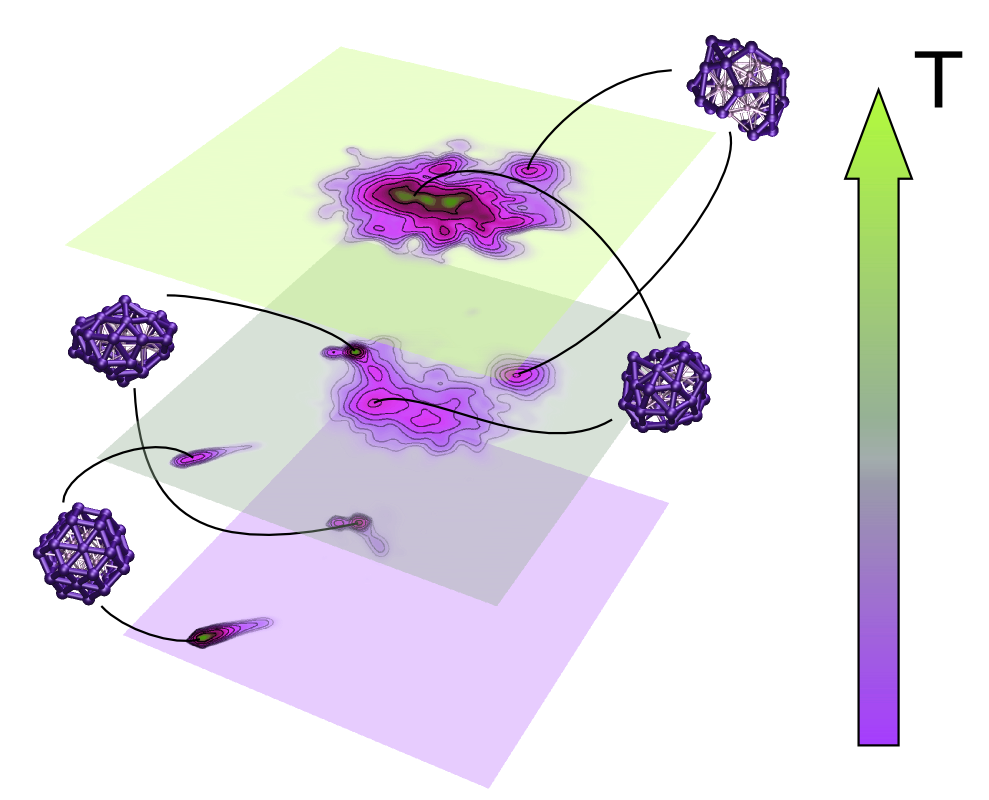}
    \caption{Figure showing the free energy surface at three different temperatures for a cluster of 38 Lennard-Jones atoms.  This particular cluster undergoes a finite-size phase transition at the temperature at which the central free energy surface in the figure above was constructed.  Furthermore, all three of the free energy surfaces above are shown as a function of a set of sketch-map coordinates that were constructed using landmark points that were taken from a trajectory at this particular temperature.  It is clear from this figure that the configurations sampled at temperatures below the transition temperature are completely different to those sampled at temperatures above the transition temperature.  At temperatures close to the transition temperature, however, the system is able to sample from both of these regions of configuration space.}
    \label{fig:LJ_clusters}
\end{figure}

Figure \ref{fig:LJ_clusters} gives an example that shows how sketch-map can be used to understand how changes in the conditions affect the free energy landscape.  This figure shows the free energy surfaces for a 38-atom cluster of Lennard Jonesium at three different temperatures \cite{smap-robust}.  This particular cluster is interesting because it has a energy landscape with a double funnel and because it therefore undergoes a finite-size phase transition from an ordered form to a disordered form \cite{lj38,lj38_pt1,lj38_pt2,lj38-path}.  The free energy surfaces that are shown in figure \ref{fig:LJ_clusters} are thus for a temperature below the phase boundary, at a temperature close to the phase boundary and at at temperature that is above the phase boundary.  The same set of sketch-map coordinates were used to construct each of these three free energy surfaces.  It is therefore possible to perform a direct comparison between them and to consequently work out what parts of configuration space this particular system explores at each temperature.  It is perhaps not surprising to note that the system is trapped in one of two small regions of configuration space at low temperature.  Furthermore, these two regions correspond to the two ordered structures that this system adopts at low temperatures.  As the temperature is raised the system is progressively allowed to explore more and more of configuration space.  Consequently, when the system is close to the transition temperature it will sample ordered and disordered configurations.  For temperatures above the transition temperature, however, entropy plays the principle role in determining the configurations that the system samples from.  The system therefore no longer samples the ordered configurations and is instead disordered at all times.   

The results shown in figure \ref{fig:LJ_clusters} are perhaps obvious given the predictions of statistical mechanics.  As temperature is increased of course the system samples from a wider portion of configuration space.  What is pleasing about the representation that is generated using the sketch-map coordinates, however, is that one really sees that that the system is sampling a larger part of configuration space at the higher temperature.  When one uses coordinates based on physical or chemical intuition by contrast this broader sampling of phase space is not always evident in the projection of the higher-temperature trajectories.

Recent work by Ardevol \emph{et al.} \cite{albert-beta1}, has shown how the sort of analysis that was demonstrated in figure \ref{fig:LJ_clusters} can be used to understand the behavior of biomolecules.  Ardevol \emph{et al.} were interested in how mutations in the amino acid sequence affected the free energy surface for the C-terminal fragment of the immunoglobulin binding domain B1 of protein G of Streptococcus (amino acid sequence Ace-GEWTYDDATKTFTVTE-NMe).  To answer this question they thus constructed a representation of a parallel tempering + metadynamics trajectory \cite{PTmetad} for the wild type protein using the sketch-map algorithm.  They then projected the wild-type trajectory using these coordinates as well as similar trajectories that were generated for each of the mutants under investigation.  They were then able to plot the free energy surfaces for the wild type and for the mutant proteins side by side and to do a point-by-point comparison between them.  From this sort of analysis they were thus able to determine what features were stabilized by the mutation and what features were destabilized by the mutation.  Furthermore, by looking at the chemical structure of the wild type and mutant they were then able to determine which chemical features were responsible for the differences in the free energy landscape.

This idea of using the sketch-map representation for one trajectory to analyse a second different trajectory can be taken a step further once you recognize that the data you analyze using these machine learning algorithms does not have to come from a molecular dynamics trajectory.  You can, for example, use a dimensionality reduction algorithm to construct a low dimensional representation for the structures in databases such as the Protein Data Bank (www.rcsb.org) \cite{pdb1,pdb2,pdb3}.  An analysis such as this can provide you with a set of generalized collective variables that can then be used to study trajectories for a range of biomolecules.  An idea similar to this one was recently used by Ardevol \emph{et al.} \cite{albert-beta2}.  They took every 16-residue fragment contained in the 7846 NMR-solved structures deposited in the PDB data bank and constructed a sketch-map representation of these structures.  They then used this projection to analyze a parallel tempering trajectory for the C-terminal fragment of the immunoglobulin binding domain B1 of protein G of Streptococcus.  They showed that the general coordinates that were constructed using data from the protein data bank were as good at discriminating between the various structures that were adopted during the trajectory as sketch-map coordinates that were constructed using the trajectory data directly.  This result suggests that it might be possible to use generic coordinates using some particularly representative data set to analyze a range of different protein systems.  These generic coordinates would provide a single common basis that would be useful when it comes to comparing the behaviors of these various different proteins.      

\subsection{Enhanced sampling}

In the previous section we showed how the sketch-map algorithm has been used to visualize trajectory data.  What was not really discussed in great detail was the way in which the analyzed trajectories were generated.  This question of how you generate trajectories to analyze is critical, however, as any projection that you generate can only ever be as informative as the data that was used to generate it.  If the trajectory that is input into the dimensionality reduction did not explore all the energetically accessible parts of configuration space any projection of this data that is generated will only provide a partial insight into the behavior of the protein.  To resolve this impasse a number of researchers have suggested using the projections that are extracted using these algorithms to enhance the sampling of phase space in one of two ways:

\begin{enumerate}
    \item A short MD trajectory is generated and then analyzed using a dimensionality reduction algorithm \cite{smap-extend}.  When the projected data is visualized some regions of the low dimensional space are found to be densely sampled, while other parts are found to be sampled more sparsely.   To broaden the sampling the researchers thus seed new trajectories using configurations taken from these sparsely sampled regions.
    
    \item The low dimensional projections obtained using a dimensionality reduction algorithm is used as a collective variable (CV) and a simulation bias that is a function of this variable is constructed using techniques such as metadynamics \cite{laio-parr02pnas}.  This simulation bias forces the system to more fully explore configuration space. 
\end{enumerate}

The first of the two methods described above is relatively self explanatory and we will thus not dwell on it much further.  Similarly, if a linear dimensionality reduction algorithm such as PCA is used it is straightforward to use this as a CV for metadynamics \cite{pca-spiwok,gervasio-pca}.  After all the CV in this case is just a linear combination of some, usually easy to calculate, set of physical parameters.  What is more challenging in this second case is if the CV is some non-linear combination of these physical parameters that is generated via a method such as sketch-map \cite{IsoMetad}.  This business of how to run enhanced sampling calculations using sketch-map as the CV will thus be the focus in the remainder of this section.  

\begin{figure}
    \centering
    \includegraphics[width=\textwidth]{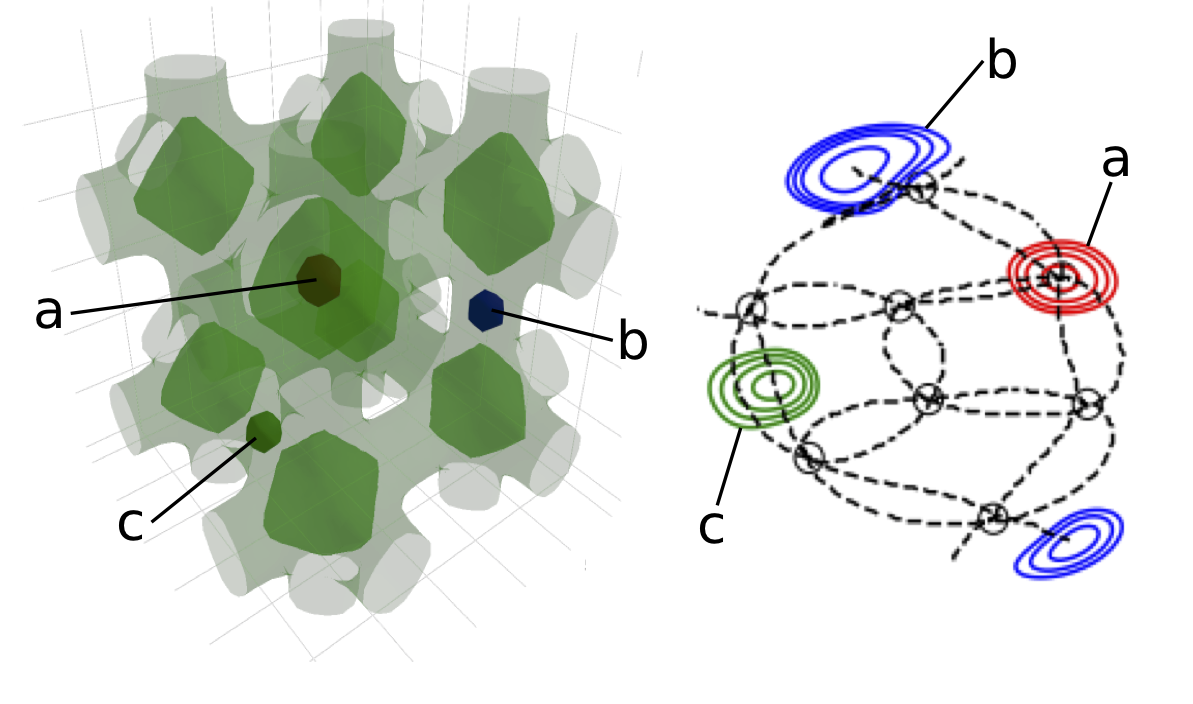}
    \caption{Figure illustrating the problems associated with using sketch-map coordinates as CVs for enhanced sampling.  To illustrate these problems we have used the energy landscape that was introduced in figure \ref{fig:3dbox-data} once more and two isosurfaces in this energy landscape are shown in the left panel above.  The right panel shows a representation of the sketch-map projection for this landscape.  The projections of each of the basins are shown using a circle, while the dashed lines are used to indicate how the transition pathways between the basins are projected.  The value of equation \ref{eqn:out-of-sample} has been evaluated on a grid in the low dimensional space for the three points  on the energy landscape labelled a, b and c.  Isocontours in these functions are shown in the right panel.  As you can see while there is a single minimum in this function and thus a single location where it is reasonable to project points a and c, there is a double minimum when this function is evaluated for point b.  It is thus difficult to know where to place the projection of this coordinate and small changes in the position of the point in the high-dimensional space can lead to large changes in the position of the projection.    }
    \label{fig:fields-problem}
\end{figure}

For sketch-map, unlike some of the other algorithms discussed in the previous sections, it is relatively simple to generate an out-of-sample projection, $\mathbf{s}$ for an an arbitrary high dimensional configuration, $\mathbf{X}$, by minimizing the following function:
\begin{equation}
\chi^2(\mathbf{s} |\mathbf{X}) = \sum_{i=1}^N w_i \left\{ F[D(\mathbf{X},\mathbf{X}_i)] - f[ d(\mathbf{s},\mathbf{s}_i)] \right\}^2
\label{eqn:out-of-sample}
\end{equation}
The sum here runs over the set of landmark points that were used to generate the initial projection.  $\mathbf{X}_i$, $\mathbf{s}_i$ and $w_i$ are the high-dimensional coordinates, the projection and the weight of landmark configuration $i$ respectively.  $D(\mathbf{X},\mathbf{X}_i)$ and $d(\mathbf{s},\mathbf{s}_i)$ thus measure the distance between the high-dimensional coordinates of the out-of-sample point and the high dimensional coordinates of the $i$th landmark and the distance between the projection of the point and the projection of the $i$th landmark.  Furthermore, in the expression above these two distances are transformed by the sigmoid functions that were discussed in section \ref{sec:dim-red}.  This stress function is thus large for $\mathbf{s}$ values for which the transformed distances to the projections of the landmarks are very different to the transformed dissimilarities from the high-dimensional coordinates.  It is small when these two sets of transformed distances are similar, which ensures that the projected landmarks that are close to $\mathbf{s}$ are those of the landmarks that are close to $\mathbf{X}$ in the high dimensional space.  This way of constructing out-of-sample projections has been shown to be very robust \cite{smap-robust} but it is, nevertheless, not possible to use the projections generated by minimising equation \ref{eqn:out-of-sample} as a CV for metadynamics \cite{fieldcvs}.  The problem with this approach is illustrated in figure \ref{fig:fields-problem}.  Essentially, the low dimensional space in which the trajectory is projected may have a different topology to the energy landscape on which the protein moves.  Consequently, paths that appear to be discontinuous in the low-dimensional projection may in actuality be continuous in the high-dimensional space.  In other words, the value of the CV that is calculated by minimizing equation \ref{eqn:out-of-sample} can change by a significant amount even when the displacement in the atomic positions is only small.

To resolve this problem with using sketch-map coordinates as a CV for metadynamics simulations we introduced the notion of a field CV \cite{fieldcvs}.  In this technique the state of the system is represented by the following function:
$$
\phi[\mathbf{s}|\mathbf{X}(t)] = \frac{ \exp\left(-\frac{\chi^2[\mathbf{s}|\mathbf{X}(t)]}{2\sigma^2}\right)}{\int \exp\left(-\frac{\chi^2[\mathbf{s}'|\mathbf{X}(t)]}{2\sigma^2} \right) \textrm{d}\mathbf{s}'}
$$
Here $\chi^2[\mathbf{s}|\mathbf{X}(t)]$ is the stress function that is defined in equation \ref{eqn:out-of-sample}.  The high dimensional coordinates, $\mathbf{X}(t)$, for the configuration can be thought of as a set of parameters that define this probability distribution, which is calculated on a grid of points, $\mathbf{s}$, in the low-dimensional space.  The probability distributions that are defined using this formula are then used in place of the Gaussians that appear in metadynamics.  There is thus a history dependent bias of sorts in this field CV method that is simply:
$$
v(\mathbf{s},t) = \sum_{t'=0}^t w(t') \phi[\mathbf{s}|\mathbf{X}(t')]
$$
where $w(t')$ is analogous to the heights of the Gaussians in metadynamics.  This quantity is time dependent because we use the standard techniques of well-tempered metadynamics (see chapter IV) \cite{wt-metad} to ensure that the bias converges.

In addition to using fields in place of the Gaussians when constructing the bias another major difference between the field-cv technique and metadynamics is the manner in which the history-dependent bias acts upon the system.  Rather than calculating the value of the history-dependent bias for the instantaneous value of the CV the field CV method calculates the instantaneous bias by performing the following integral:
$$
V[\mathbf{X}(t)] = \int \phi[\mathbf{s}|\mathbf{X}(t)]v(\mathbf{s},t) \textrm{d}\mathbf{s}
$$
As shown in figure \ref{fig:fields-problem} calculating the instantaneous bias using this equation resolves the issues associated with continuous paths in the high dimensional being projected as discontinuous paths in the low-dimensional space.  In essence the system now deposits bias in all the part of the low dimensional space where it would be reasonable to project the configurations.  Furthermore, at any given time the system feels the bias  that has been deposited in all the points where it would be reasonable to project the configuration.  

Simulations that demonstrate that the field CV method that has been outlined in the previous paragraphs can be used to enhance the sampling in model systems have been performed \cite{fieldcvs}.  The method shows considerable promise but it is currently computationally expensive to run and thus has only been rarely used.  It is, however, an interesting approach and one that should be investigated further in the future.

\section{Conclusions}

The chapter has discussed how machine learning algorithms can be used to visualize molecular dynamics trajectories and to enhance sampling.  There has been a veritable explosion of interest in using these techniques to understand simulation data in the past few years and as such any presentation on this topic will probably barely scratch the surface of the literature.  What we hope that we have provided in the preceding pages is an easy-to-digest-but-far-from-exhaustive introduction to some of the ideas that are being used.  In this final section we would like to finish by briefly discussing some interesting recent directions in which we believe the field is moving.

Throughout this chapter we have asserted that these methods should be used to complement chemical and physical understanding and not to replace it.  With this in mind an interesting recent development is the so called PAMM methodology \cite{pamm1,pamm2}, which uses Bayesian statistics to determine whether the arrangement of the atoms in a particular configuration resembles the canonical definition of a molecular motif such as a hydrogen bond or alpha helix.  This method is appealing as physical intuition and machine learning are used in tandem.  Finding appropriate fingerprint vectors to encode our physical understanding remains a challenge, however, and some have argued that we should instead use more generic representations to describe the arrangement of the atoms \cite{smap-soap-1,smap-soap-3}. 

A second interesting recent direction has involved applying the deep learning techniques that have proved so successful in a range of fields to biophysical problems.  In particular, a number of recent articles have used autoencoder neural networks to construct collective coordinates that can be used both to analyze molecular dynamics trajectories and as a collective variable for metadynamics simulations \cite{ferg-auto-encoders,pande-auto-encoders}.

Finally, most of the algorithms discussed in this chapter do not consider the order that the frames are visited in within the trajectory.  Consequently, any projections that are constructed reproduce the spatial relationships between the frames in the input trajectories rather than the temporal relationships.  Recent developments in Markov State Modelling \cite{msms,clementi-and-noe} and the development of techniques for extracting rate constants from enhanced sampling calculations \cite{pratyush-reweight} perhaps provide ways of generating low-dimensional projections that incorporate information on the temporal information in the trajectory \cite{more-clementi-and-noe,sgoop,pande-auto-encoders,tica-metad}.  In other words, these new techniques generate low dimensional coordinates that describe the directions in which the system diffuses slowly by analyzing transition probability matrices directly.  This form of analysis is an exciting development as the projections that emerge would provide real insight into the slow-degrees of freedom and hence the reaction mechanisms. 

\section{Notes}
\begin{enumerate}
\item
\label{note:distance-matrix}
We can write out all the matrix elements for a $3 \times 3$ matrix of distances using equation \ref{eqn:pythagoras} and thus see that equation \ref{eqn:dissims-matrices} holds:
$$
\begin{aligned}
\left[ \begin{array}{ccc}
0 & d_{12}^2 & d_{13}^2 \\
d_{12}^2 & 0 & d_{23}^2 \\
d_{13}^2 & d_{23}^2 & 0 
\end{array}\right] 
 = & 
\sum_\alpha \left[ \begin{array}{ccc}
(X^{(1)}_\alpha)^2 & (X^{(1)}_\alpha)^2 & (X^{(1)}_\alpha)^2 \\
(X^{(2)}_\alpha)^2 & (X^{(2)}_\alpha)^2 & (X^{(2)}_\alpha)^2 \\
(X^{(3)}_\alpha)^2 & (X^{(3)}_\alpha)^2 & (X^{(3)}_\alpha)^2 \\
\end{array}\right] 
  + \sum_\alpha \left[ \begin{array}{ccc}
(X^{(1)}_\alpha)^2 & (X^{(2)}_\alpha)^2 & (X^{(3)}_\alpha)^2 \\
(X^{(1)}_\alpha)^2 & (X^{(2)}_\alpha)^2 & (X^{(3)}_\alpha)^2 \\
(X^{(1)}_\alpha)^2 & (X^{(2)}_\alpha)^2 & (X^{(3)}_\alpha)^2 \\
\end{array}\right] \\
& - 2 \sum_\alpha \left[ \begin{array}{ccc}
X^{(1)}_\alpha X^{(1)}_\alpha & X^{(1)}_\alpha X^{(2)}_\alpha & X^{(1)}_\alpha X^{(3)}_\alpha \\
X^{(2)}_\alpha X^{(1)}_\alpha & X^{(2)}_\alpha X^{(2)}_\alpha & X^{(2)}_\alpha X^{(3)}_\alpha \\
X^{(3)}_\alpha X^{(1)}_\alpha & X^{(3)}_\alpha X^{(2)}_\alpha & X^{(3)}_\alpha X^{(3)}_\alpha
\end{array}\right]
\end{aligned}
$$

\item
\label{note:centering}
The centering matrix, $\mathbf{J}$, that was introduced in equation \ref{eqn:centering-matrix} has the useful property that $\mathbf{1}^T \mathbf{J} = \mathbf{J}\mathbf{1}=\mathbf{0}$, where $\mathbf{0}$ is a matrix of zeros.  We thus find if we multiply the matrix $\mathbf{D}$ that was introduced in equation \ref{eqn:dissims-matrices} from the front and the back by $-\frac{1}{2} \mathbf{J}$ that:
$$
-\frac{1}{2} \mathbf{J}\mathbf{D}\mathbf{J} = -\frac{1}{2} \mathbf{J} \mathbf{c}\mathbf{1}^T\mathbf{J} - \frac{1}{2} \mathbf{J}\mathbf{c}^T \mathbf{1}\mathbf{J} + \mathbf{J} \mathbf{K}\mathbf{J} = \mathbf{J} \mathbf{K}\mathbf{J}
$$
Furthermore, by substituting in our expression for $\mathbf{J}$ we find:
$$
-\frac{1}{2} \mathbf{J}\mathbf{D}\mathbf{J} = \mathbf{K} -\frac{1}{M^2} \mathbf{1} \mathbf{1}^T \mathbf{K} \mathbf{1} \mathbf{1}^T
$$

Every element of $\mathbf{1} \mathbf{1}^T \mathbf{K} \mathbf{1} \mathbf{1}^T$ is equal to the sum of the elements of $\mathbf{K}$ so the above manipulations demonstrate that the centered matrix of distances, $-\frac{1}{2} \mathbf{J}\mathbf{D}\mathbf{J}$, is equal to the Gram matrix of kernels modulo an additive constant.

\item
\label{note:diffusion-thing}
It is possible to introduce further sophistication into Laplacian Eigenmaps by introducing a diffusion kernel.  When this modification is used the distances between each $\mathbf{x}_i$ and each of its $k$ nearest points, $\mathbf{y}_j$ is transformed using the following isotropic diffusion kernel:
\begin{equation}
\mathbf{P}_{ij} = \mathbf{P}(\mathbf{x}_i,\mathbf{y}_j ) = \exp\left(-\frac{|\mathbf{x}-\mathbf{y}|^2}{\sigma} \right)
\label{eqn:kernelf}
\end{equation}
where $\sigma$ is a hyperparameter.  This diffusion kernel is at the heart of diffusion maps, which works by calculating this quantity for each pair of input data points without first computing the $k$ nearest points or the pairs of data point that are within a certain cutoff.

\item
\label{note:diffusion-maps}
In diffusion maps a weighted graph $\mathbf{P}$ is calculated using equation \ref{eqn:kernelf}.  This graph is then transformed using:
$$
\widehat{\mathbf{P}}_{ij} = \frac{\mathbf{P}_{ij} }{ \sqrt{ \mathbf{D}_{ii} \mathbf{D}_{jj} } } 
$$
to give a matrix $\widehat{\mathbf{P}}$ that is equal to the identity minus the symmetric-normalized Laplacian of the graph $\mathbf{P}$.  From this matrix we then compute $\widehat{\mathbf{D}}$ using:
$$
\widehat{\mathbf{D}}_{ij} = 
\begin{cases}
\sum_{j\ne i} \widehat{\mathbf{P}}_{ij} & \textrm{if} \quad i=j \\
0 & \textrm{otherwise}
\end{cases}
$$
we then obtain an $M \times N$ matrix, $\widehat{\mathbf{X}}$, with low dimensional projections for the $M$ input points in its rows by diagonalizing $\widehat{\mathbf{D}}^{-\frac{1}{2}} \widehat{\mathbf{P}} \widehat{\mathbf{D}}^{-\frac{1}{2}}$, discarding the largest eigenvalue and its corresponding eigenvector and by then taking the eigenvectors corresponding to the $N$ largest eigenvalues that remain and placing them in the rows of $\widehat{\mathbf{X}}$.

\item
\label{note:similarity}
The eigenvectors of the matrix that is diagonalized in diffusion maps, $\widehat{\mathbf{D}}^{-\frac{1}{2}} \widehat{\mathbf{P}} \widehat{\mathbf{D}}^{-\frac{1}{2}}$, are related by a relatively simple transformation to the eigenvectors of $\widehat{\mathbf{D}}^{-1} \widehat{\mathbf{P}}$. This matrix is similar to the matrix that appeared in equation \ref{eqn:laplacian} and that is diagonalized in Laplacian Eigenmaps. 

\item
\label{note:chapman}
The Chapman-Kolmorov relation tells us that if we are given a one step transition probability matrix for a Markov chain, $\mathbf{P}$ we can extract the $t$-step transition probability matrix by raising $\mathbf{P}$ to the $t$th power.  It is well established, however, that we can write the $t$th power of this transition matrix as:
\begin{equation}
\mathbf{M}^t = \mathbf{V} \mathbf{\Lambda}^t \mathbf{V}^{-1} 
\label{eqn:diag}
\end{equation}
where $\mathbf{V}$ is a matrix containing the eigenvectors of $\mathbf{M}$ in its columns and where $\mathbf{\Lambda}$ is a diagonal matrix that contains the eigenvalues of $\mathbf{M}$.  Calculating the $t$th power of a diagonal matrix involves simply raising each element to the power $t$.  Applying this procedure to equation \ref{eqn:diag} will therefore widen the gap between the largest and smallest eigenvalues.  Furthermore, when equation \ref{eqn:diag} is used to recompose $\mathbf{M}^t$ each of the exponentiated eigenvalues is only multiplied by its corresponding eigenvector.  We thus find that, when $t$ is large, the matrix, $\mathbf{M}^t$ that we would construct by entering only the largest few eigenvalues and their corresponding eigenvectors into equation \ref{eqn:diag} is very similar to the matrix that we would have obtained had we used all the eigenvalues and eigenvectors when evaluating equation \ref{eqn:diag}.  

\item
\label{note:laplacian}
The matrix $\mathbf{P}$ that is diagonalized in diffusion maps is related to the symmetric-graph Laplacian, $\widehat{\mathbf{L}} = \mathbf{I} - \widehat{\mathbf{P}}$.  Graph Laplacians of this sort appear in Laplacian Eigenmaps.  Furthermore, the eigenvectors of $\widehat{\mathbf{L}}$ are identical to those of $\widehat{\mathbf{P}}$.  In addition, the eigenvalues, $\lambda$, of $\widehat{\mathbf{L}}$ are related to those of $\widehat{\mathbf{P}}$ by $1-\lambda$.  Consequently, because $\widehat{\mathbf{P}}$ is a positive matrix with eigenvalues that are all positive, the eigenvectors that correspond to the largest eigenvalues of $\widehat{\mathbf{P}}$ will be equal to the eigenvectors that correspond with the smallest eigenvalues of $\widehat{\mathbf{L}}$.  This is why one takes the eigenvectors corresponding to the smallest eigenvalues when using Laplacian eigenmaps and the eigenvectors corresponding to the largest eigenvalues when using diffusion maps.


\end{enumerate}

\bibliographystyle{unsrt}
\bibliography{biblio}

\end{document}